\documentclass[
    reprint,
    nofootinbib,
    aps,
    prd,
    amsmath,
    floatfix,
    ]{revtex4-2}
\pdfoutput=1 

\usepackage[T1]{fontenc} 
\usepackage{bm}
\usepackage{graphicx}
\usepackage{hyperref}
\usepackage{xcolor}
\usepackage{multirow}
\usepackage{hhline}
\usepackage{enumerate}

\usepackage{array}

\definecolor{ShamrockGreen}{rgb}{0.0, 0.62, 0.38}
\definecolor{LightBlue}{rgb}{0.42, 0.82, 1.90}

\makeatletter
\def\l@subsubsection#1#2{}
\makeatother

\usepackage[toc,page]{appendix}

\newcommand{\nocontentsline}[3]{}
\newcommand{\tocless}[2]{\bgroup\let\addcontentsline=\nocontentsline#1{#2}\egroup}

\begin{document}

\title{Constraining Dark Photons with Self-consistent Simulations of Globular Cluster Stars}

\author{Matthew J. Dolan}
\email{matthew.dolan@unimelb.edu.au}
\author{Frederick J. Hiskens}
\email{fhiskens@student.unimelb.edu.au}
\author{Raymond R. Volkas}
\email{raymondv@unimelb.edu.au}
\affiliation{ARC Centre of Excellence for Dark Matter Particle Physics, School of Physics, The University of Melbourne, Victoria 3010, Australia}

\date{\today}

\begin{abstract}
We revisit stellar constraints on dark photons. We undertake dynamical stellar evolution simulations which incorporate the resonant and off-resonant production of transverse and longitudinal dark photons. We compare our results with observables derived from measurements of globular cluster populations, obtaining new constraints based on the luminosity of the tip of the red-giant branch (RGB), the ratio of populations of RGB to horizontal branch (HB) stars (the $R$-parameter), and the ratio of asymptotic giant branch to HB stars (the $R_2$-parameter). We find that previous bounds derived from static stellar models do not capture the effects of the resonant production of light dark photons leading to overly conservative constraints, and that they over-estimate the effects of heavier dark photons on the RGB-tip luminosity. This leads to differences in the constraints of up to an order of magnitude in the kinetic mixing parameter.
\end{abstract}

\maketitle
\flushbottom

\tableofcontents

\section{Introduction}

Many extensions of the Standard Model (SM) of particle physics include dark sectors, which may involve  the introduction of a new $U(1)$ gauge group with an associated gauge boson, the \textit{dark photon} (DP). Dark photons interact with SM particles via kinetic mixing with the visible photon, which at low energies is given by the following Lagrangian:
\begin{equation}
    \label{Eq: Mixing lagrangian}
    \mathcal{L}=-\frac{1}{4}F_{\mu\nu}F^{\mu\nu}-\frac{1}{4}V_{\mu\nu}V^{\mu\nu}+\frac{m_{\mathrm{DP}}^2}{2}V_{\mu}V^{\mu}-\frac{\chi}{2}F_{\mu\nu}V^{\mu\nu},
\end{equation}
where $F_{\mu\nu}$ is the standard electromagnetic field strength tensor, $V_{\mu\nu}$ is its dark counterpart, $V_{\mu}$ is the dark photon field and $\chi$ is the kinetic mixing parameter.

Unlike their visible counterparts, dark photons may have a non-zero physical mass, $m_{\mathrm{DP}}$, which can be generated through either the Stueckelberg or a dark Higgs-type mechanism. Should they be non-thermally produced in the early universe, dark photons can constitute a viable dark matter candidate \cite{Arias:2012az}. Alternatively, they may serve as a portal between distinct visible and dark sectors.

\begin{figure*}[t]
    \centering
    \includegraphics[width = 0.8\textwidth]{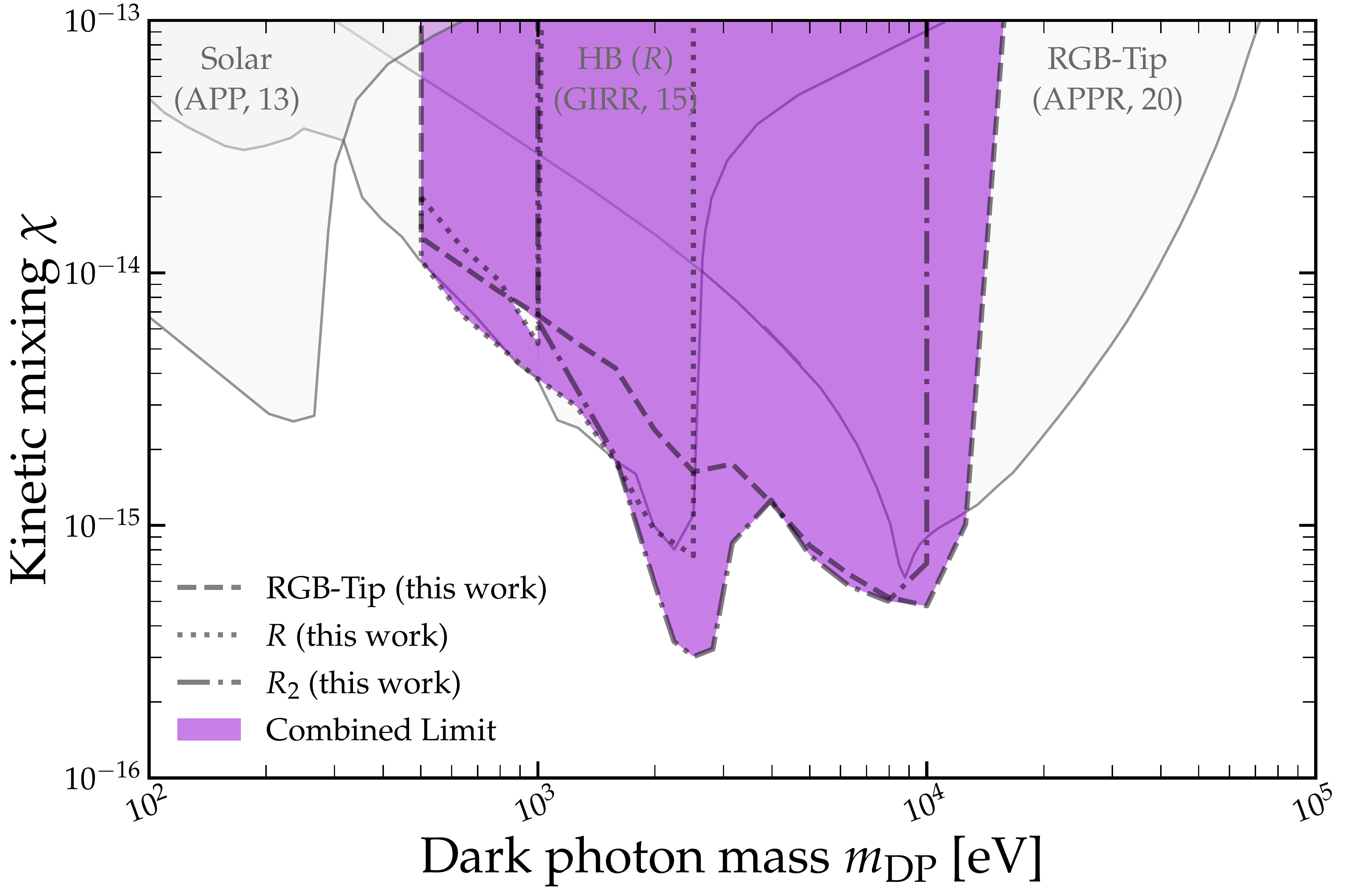}
    \caption{Dark photon parameter space for $100$~eV $\leq m_{\mathrm{DP}}\leq 100$~keV, showing our combined limit in purple. Our individual bounds from the RGB-tip luminosity, $R$ and $R_2$ are indicated by the dashed, dotted and dot-dashed lines respectively. These limits are cut off at $m_{\rm{DP}}=501$~eV, below which dark photon energy-loss substantially disrupts main sequence evolution. We also show constraints from the sun (APP, 13) \cite{An:2013yfc}, and previous bounds from the $R$-parameter (GIRR, 15) \cite{2016JCAP...05..057G} and the RGB-tip luminosity (APPR, 20) \cite{An:2020bxd}. The grey shaded regions are excluded according to these references. }
    \label{fig: DP parameter space}
\end{figure*}

Light and weakly interacting dark photons, which naturally arise in, for example, string compactifications \cite{Goodsell:2009xc}, are subject to the suite of constraints from detectors, cosmology and astrophysics one usually applies to weakly interacting slim particles (WISPs), such as axions and axion-like particles (ALPs). Included in this is their potential to be produced in, and drain energy from, the deep interior of stars, which can lead to stellar cooling constraints via the usual energy-loss argument \cite{Raffelt:1996wa}.

A number of such stellar cooling constraints exist in the literature, including those derived from massive stars~\cite{Sieverding:2021jfa}, the sun~\cite{An:2013yfc, Redondo:2013lna, Vinyoles2015:10015V} and observables related to globular clusters ~\cite{Redondo:2013lna, An:2013yfc, AN2015331, 2016JCAP...05..057G, An:2020bxd} -- including the ratio of horizontal branch (HB) to red-giant branch (RGB) stars (the $R$-parameter) and the RGB-tip luminosity. We show these bounds in Fig.~\ref{fig: DP parameter space} in grey, which includes the solar, $R$ and RGB-tip limits of Refs.~\cite{An:2013yfc,2016JCAP...05..057G,An:2020bxd}. Similar bounds were obtained by Ref.~\cite{Redondo:2013lna}.

All of the previous bounds presented in grey in Fig.~\ref{fig: DP parameter space} have been constructed by:\ (i) converting existing stellar bounds on axions to numerical upper limits on either the energy-loss rate or total novel luminosity, and (ii) demanding that the dark photon energy-loss rate per unit mass be smaller than this when averaged across a \textit{static} stellar profile or integrated over the entire star. Their reliability can therefore only be guaranteed if energy-loss to dark photons affects stellar evolution in a similar way to the axion case. As we show in this work, this is not always true. In light of this, \textit{the goal of this paper is to use dynamic and self-consistent stellar evolution simulations to set updated limits on the $m_{\mathrm{DP}}$-$\chi$ plane.}

For certain regions of parameter space, the production of transverse dark photons in stars occurs resonantly in stellar regions with plasma frequency equal to the dark photon mass. If this happens, the resonant process dominates the total energy-loss. Unusually for novel energy-loss mechanisms, the resonant production region (RPR) may be off centre and, crucially, moves throughout the evolution of the star. Consequently, bounds derived from static models either fail to capture the full influence of this source of energy-loss or, for some dark photon masses, overestimate its effects on the stellar parameters from which the constraints are derived.

We have also found that the existence of such an RPR in a star can lead to interplay between the novel energy-loss and the convective structure of the star. The ratio of asymptotic giant branch (AGB) to HB stars in globular clusters (the $R_2$-parameter), which we recently used to set a new constraint on the axion-photon coupling \cite{Dolan:2022kul}, is a particularly sensitive probe of this phenomenon. 

Our combined limit is shown in purple in Fig.~\ref{fig: DP parameter space}, with the individual constraints from the RGB-tip luminosity, $R$ and $R_2$ indicated by the dashed, dotted and dot-dashed lines respectively. These have been cutoff at $m_{\rm{DP}}=501$~eV, as stars including energy-loss to dark photon lighter than this experience significantly disrupted main sequences, hindering our ability to simulate beyond this phase.

Our results, which supersede the RGB-tip and $R$ limits of Refs.~\cite{2016JCAP...05..057G, An:2013yfc}, shows an improvement of up to an order of magnitude in the mass range between 2~keV and 10~keV which originates from two different sources. The first, which corresponds to the stalactite-like protuberance reaching down to $\chi\approx3\times10^{-16}$ at $m_{\rm{DP}}\approx2.5$~keV, arises due to interplay between dark photon energy-loss and the convective structure of horizontal branch stars, which can be ruled out by $R_2$. The second region of improvement (in the mass range $3$~keV$\lesssim m_{\rm{DP}}\lesssim8$~keV) occurs due to resonant dark photon production during the progression of the RGB and AGB phases, which is not captured by static models of the former. 

Furthermore, we find that the constraining power of the RGB-tip luminosity has been over-estimated in the parameter-space above $m_{\rm{DP}}\gtrsim12.6$~keV, potentially due to the use of static models as well as the neglected impact of electron degeneracy and screening on the adopted energy-loss rates. Stellar constraints in this region should therefore relax. Finally, we note the existence of a small unconstrained triangular region at $m_{\rm{DP}}\sim1$~keV, which arises due to the competing influence of dark photon energy-loss on the RGB and HB phases.

The structure of this paper is as follows. In Sec.~\ref{sec: sec2} we describe the process by which energy-loss to dark photons occurs. Following this, we discuss the existing RGB-tip constraints on dark photons in more detail and present our own limit in Sec.~\ref{sec: RGB tip luminosity}. In Sec. \ref{sec: star cluster counts}, we do the same for the bounds from $R$ and $R_2$, giving particular attention to the interplay between energy-loss to dark photons and convection described above, before summarising and concluding in Sec.~\ref{sec: conclusion}.

\section{Dark photon production in stars}
\label{sec: sec2}
The production of dark photons in globular cluster stars is typically discussed in terms of photon-dark photon mixing in conjunction with plasma photon production and absorption \cite{Redondo:2008aa, Redondo:2013lna, An:2013yfc}. A stellar plasma can support transverse and longitudinal photons, both of which oscillate into their corresponding dark photon polarisations. If the dark photons have sufficiently long lifetimes they can freely escape the star, contributing to energy-loss from the stellar interior. This affects stellar structure by reducing the local energy production rate per unit mass $\epsilon$, an input to the stellar structure equations \cite{Raffelt:1996wa}.

\subsection{Energy-loss rates}
\label{sec: energy-loss}

We can express the energy-loss rates per unit mass to longitudinal (L) and transverse (T) dark photons as an integral over dark photon energy $\omega$,
    \begin{equation}
        \label{eq: energy-loss}
        \epsilon_{\rm{L,T}}=\frac{g_{\rm{L,T}}}{2\pi^2\rho}\int_{m_{\mathrm{DP}}}^{\infty}d\omega\, \omega^2\sqrt{\omega^2-m_{\mathrm{DP}}^2}\, \Gamma_{\mathrm{Prod}}^{(\rm{L,T})},
    \end{equation}
where $\rho$ is the local stellar density, $g_{\rm{L,T}}$ are the degeneracy factors for longitudinal ($g_{\rm{L}}=1$) and transverse ($g_{\rm{T}}=2$) modes and $\Gamma_{\mathrm{Prod}}^{(\mathrm{L,T})}$ are the longitudinal and transverse dark photon production rates. The total energy-loss $\epsilon_{\mathrm{DP}}$ is given as a sum over both modes\footnote{Note that Refs \cite{Sieverding:2021jfa}, \cite{Chang:2016ntp} and \cite{Rrapaj:2019eam} include energy-loss to dark photons produced via electron-positron pair annihilation, which is only relevant in the later evolutionary phases of stars with $M_{\rm{init}}\gtrsim8M_{\odot}$ (i.e. carbon burning and beyond), which are hot enough to support sizeable positron populations.}.

In a non-relativistic plasma, expressions for $\Gamma_{\mathrm{Prod}}^{(\mathrm{L,T})}$ are given by \cite{Redondo:2013lna}
    \begin{equation}
    \label{eq: L production}
        \Gamma_{\mathrm{Prod}}^{(\mathrm{L})}=\frac{\chi^2m_{\mathrm{DP}}^2}{e^{\omega/T}-1}\frac{\omega^2\Gamma_{\rm{L}}}{(\omega^2-\omega_{\mathrm{pl}}^2)^2+(\omega\Gamma_{\rm{L}})^2}
    \end{equation}
    and 
    \begin{equation}
        \label{eq: T production}
        \Gamma_{\mathrm{Prod}}^{(\mathrm{T})}=\frac{\chi^2m_{\mathrm{DP}}^4}{e^{\omega/T}-1}\frac{\Gamma_{\rm{T}}}{(m_{\mathrm{DP}}^2-\omega_{\mathrm{pl}}^2)^2+(\omega\Gamma_{\rm{T}})^2}
    \end{equation}
respectively. Here $\omega^2_{\mathrm{pl}}=4\pi\alpha n_e/m_e$ and $T$ are the local stellar plasma frequency and temperature,  $n_e$ is the electron number density, and $\Gamma_{\rm{L,T}}$ are the longitudinal and transverse plasmon damping rates.

Note that equations \ref{eq: L production} and \ref{eq: T production} differ in both their scaling with $m_{\mathrm{DP}}$ and their denominators. These originate from the different dispersion relations obeyed by the respective photon modes. 

\subsection{Resonant emission}

Energy-loss to both dark photon modes can proceed resonantly. $\Gamma_{\mathrm{Prod}}^{(\mathrm{L})}$ peaks for dark photon energies approximately equal to $\omega_{\mathrm{pl}}$. As $\Gamma_{\rm{L}}\ll\omega_{\mathrm{pl}}$ in all scenarios we consider, this production rate is well-approximated by a delta function \cite{Redondo:2013lna}. Consequently, longitudinal dark photons will be resonantly produced in all stellar regions with energies equal to the plasma frequency of that region, unless $m_{\mathrm{DP}}>\omega_{\mathrm{pl}}$. The energy-loss rate per unit mass associated with this is \cite{Redondo:2013lna}
\begin{equation}
    \label{eq: eps L}
    \epsilon_{\rm{L}}=\frac{\chi^2m_{\mathrm{DP}}^2}{4\pi\rho}\frac{\omega_{\mathrm{pl}}^2\sqrt{\omega_{\mathrm{pl}}^2-m_{\mathrm{DP}}^2}}{e^{\omega_{\mathrm{pl}}/T}-1}.
\end{equation}
This is the equation we implement in our stellar simulations to account for energy-loss to longitudinal dark photons.

Unlike that of their longitudinal counterparts, transverse dark photon resonant production is restricted to stellar regions with $\omega_{\mathrm{pl}}\approx m_{\mathrm{DP}}$. Consequently, it is only relevant in a given star if $m_{\mathrm{DP}}$ lies within its range of plasma frequencies. Furthermore, should such an RPR exist for a dark photon of a given mass, it may be off-centre and will move throughout the evolution of the host star. This scenario is very atypical compared with energy-loss due to axions or axion-like particles and leads to interesting and non-standard results in simulations.

Interestingly, Eqs.~\ref{eq: energy-loss} and \ref{eq: T production} imply that the height of the resonant peak in $\epsilon_{\rm{T}}$ scales inversely with $\Gamma_{\rm{T}}$, while its width scales linearly with $\Gamma_{\rm{T}}$. Consequently, the total novel energy-loss within a star does not depend sensitively on the exact value of $\Gamma_{\rm{T}}$ so long as it hosts an RPR.

\subsection{Off-resonant production}
\label{sec: off-resonant production}
As transverse dark photons are only produced resonantly in a small region within a given star (if at all), an explicit formula for $\Gamma_{\rm{T}}$ is needed.

The deep interior of the globular cluster stars we consider is non-relativistic and completely ionised with varying degrees of electron degeneracy. In these conditions, the primary contributions to $\Gamma_{\rm{T}}$ come from inverse Bremsstrahlung (free-free absorption) and Thomson scattering, while free-bound, bound-bound and electron-electron free-free absorption are of less importance. Consequently, $\Gamma_{\rm{T}}$ is given by \cite{Redondo:2008aa, Redondo:2013lna}
\begin{equation}
\label{eq: GammaT}
\begin{split}
\centering
    \Gamma_{\rm{T}}=\frac{16\pi^2\alpha^3}{3m_e^2}\sqrt{\frac{2\pi m_e}{3T}}&n_e\frac{1-e^{-\omega/T}}{\omega^3}\sum_jZ_j^2n_j\Bar{g}_{ff,j} \\ &+\frac{8\pi\alpha^2n_e}{3m_e^2},
\end{split}
\end{equation}
where $m_e$ is the electron mass, $\alpha$ is the electromagnetic fine-structure constant and $n_j$ ($n_e$) is the number density of particles with charge $Z_j$ (electrons). 

In Eq.~\ref{eq: GammaT}, $\Bar{g}_{ff, j}$ are the averaged free-free Gaunt factors for nuclear species $j$, which provide corrections to the semi-classical free-free cross section of Kramers from, for example, particle correlations (including collective effects), plasma dispersion and multiple collisions, as well as plasma degeneracy and relativistic effects \cite{Iglesias_1996}.

The accurate determination of \textit{un-averaged} free-free Gaunt factors $g_{ff, j}$ has been an area of active research from their introduction in 1930 \cite{1930RSPSA.126..654G} until now (see e.g. \cite{Pradler2021}). While an exact equation (assuming a purely Coulomb potential) was provided by Sommerfeld and Maue in 1935 in terms of hypergeometric functions~\cite{SommerfeldMaue}, its direct evaluation is often too computationally cumbersome to be of practical use inastrophysical applications. This has lead to the development of a rich catalogue of analytical approximations \cite{RevModPhys.34.507} and tabular values of $g_{ff, j}$ such as \cite{KL61}, each with varying assumptions and regions of validity.

In stellar applications, rather than being concerned with free-free absorption for specific electron velocities, we are instead interested in the bulk absorptive properties of the plasma. The appropriate quantity for us is therefore the thermal average of $g_{ff,j}$ across an appropriate electron distribution $f_e$, which is precisely the factor $\Bar{g}_{ff,j}$ that appears in Eq.~\ref{eq: GammaT}. In most scenarios this is assumed to be a Maxwell-Boltzmann distribution, but the effects of plasma electron degeneracy and Pauli blocking can be included by instead performing this average over a Fermi-Dirac distribution and introducing a quantity $q$ when integrating corresponding to the probability of the final state being unoccupied \cite{RM-2580-AEC, 1987ApJS...63..661N}.

In stellar plasmas the rate of photon absorption will also be affected by ion-charge screening\footnote{Ion-ion correlations and plasma dispersion effects can also be included via the methods of Ref.~\cite{bekefi1966radiation}. These corrections, however, are often small and have historically been excluded from stellar opacity calculations \cite{Iglesias_1996}.}. This is typically included in calculations by replacing the aforementioned Coulomb potential with a Yukawa potential, with an appropriate screening scale, such as the plasma Debye-H\"uckel wavelength $\lambda_{\mathrm{D}}$ \cite{ARMSTRONG201461}.

Finally, relativistic effects also become non-negligible when the plasma temperature is above $10^8$~K \cite{Iglesias_1996}, which coincides with helium ignition in stars. The corresponding corrections to the Gaunt factors can be evaluated by use of the relativistic cross section of Bethe and Heitler~\cite{BetheHeitler}, as was done in Refs.~\cite{Iglesias_1996} and \cite{1987ApJS...63..661N}. Numerous analytical approximations and tabular values for these averaged-Gaunt factors are also available in the literature such as \cite{RM-2580-AEC,1987ApJS...63..661N}.

The discussion above clearly demonstrates the challenge of accurately calculating $\Bar{g}_{ff,j}$ for the wide range of conditions and environments encountered during stellar evolution. 
Previous stellar dark photon constraints have opted for two different approaches. In Ref.~\cite{Redondo:2008aa}, values of $\Bar{g}_{ff,j}$ were calculated from the exact Sommerfeld-Maue formula following the method of Ref.~\cite{1958MNRAS.118..241G}, which neglects contributions from electron degeneracy, screening, ion-ion correlations, relativity and dispersion, all of which provide additional corrections of 5\% or less in the solar centre for peak photon energies \cite{Iglesias_1996}.

Reference~\cite{Redondo:2013lna}, on the other hand, adopts an analytic approximation for $\Bar{g}_{ff,j}$, which includes the effects of screening,
\begin{equation}
    \label{eq: Gbar with screening}
    \Bar{g}_{ff}(u) = \frac{2\sqrt{3}}{\pi}\int_0^{\infty}dx\, x e^{-x^2}\int_{\sqrt{x^2+u}-x}^{\sqrt{x^2+u}-x}\frac{dt\, t^3}{(t^2+y^2)^2},
\end{equation}
where $u=\omega/T$, $y=k_s/\sqrt{2m_eT}$ and $k_s$ is an appropriate screening scale. If screening is ignored (i.e. $y=0$), this result agrees with Ref.~\cite{An:2013yfc}, and the expression above reduces to \cite{Redondo:2013lna}
\begin{equation}
    \label{eq: Gbar wo screening}
    \Bar{g}_{ff}(u)=\frac{\sqrt{3}}{\pi}e^{u/2}K_0(u/2)
\end{equation}
independent of $j$, where $K_0$ is a modified Bessel function of the second kind.

While Eq.~\ref{eq: Gbar wo screening} is suitably accurate in a solar environment, the same is not true for the other evolutionary phases examined in this work and Refs.~\cite{Redondo:2013lna, An:2013yfc, AN2015331, An:2020bxd}. For example, red-giant branch cores are highly degenerate ($\eta\equiv\mu_e/T\sim20$) in the very late stages of their evolution, while horizontal branch cores have temperatures on the threshold of requiring a relativistic treatment. These facts seemingly necessitate the inclusion of corrections from screening, Pauli blocking and special relativity in order to accurately evaluate $\Gamma_{\rm{T}}$.

As previously mentioned, however, the total energy-loss across the RPR, which amounts to most of the total energy-loss to dark photons experienced by the entire star, does not depend sensitively on the precise value of $\Gamma_{\rm{T}}$. In light of this, and because a full calculation of the Gaunt factors including corrections for screening, electron degeneracy and relativity is computationally cumbersome within the confines of dynamic stellar evolution simulations, we adopt the form of $\Bar{g}_{ff,j}$ given in Eq.~\ref{eq: Gbar wo screening}. 

The validity of this assumption is assessed in App.~\ref{app: systematics}, where we consider several other benchmark values of $\Bar{g}_{ff,j}$ appropriate to the stellar regions and phases relevant for each bound. Changes in the Thomson scattering cross-section, which also has multiplicative corrections from screening, relativity and electron degeneracy \cite{Iglesias_1996}, are also examined. 

Finally, we note that the insensitivity of our derived bounds to the specific treatment of the Gaunt factors is only guaranteed when an RPR is responsible for the majority of energy-loss which facilitates the constraint. This is the case for all of our new limits with the exception of that of the RGB-tip luminosity for $m_{\mathrm{DP}}\gtrsim20$~keV. Estimates for the bound in this region are presented in App.~\ref{app: systematics}.

\section{The RGB-tip luminosity}
\label{sec: RGB tip luminosity}
\begin{figure}[t]
    \centering
    \includegraphics[width=0.45\textwidth]{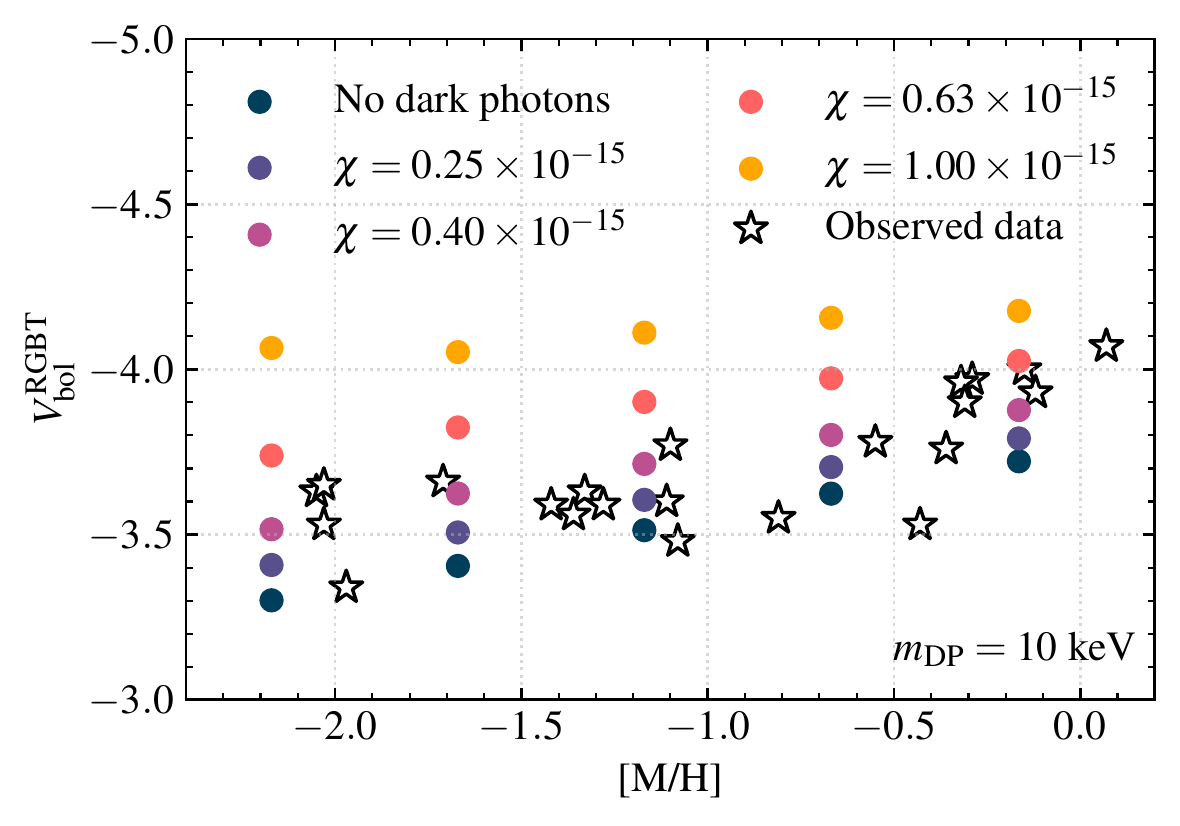}
    \caption{Predictions for the RGB-tip bolometric magnitudes of globular clusters with metallicities between $Z=0.0001$ and $Z=0.01$ generated using our edited version of \texttt{MESA}, with either no novel energy-loss or energy-loss to $10$~keV dark photons with $\chi$ ranging between $2.51\times10^{-15}$ and $10^{-15}$.}
    \label{fig: RGB-T predictions v observations}
\end{figure}

The red-giant branch is a feature in Galactic and globular cluster Hertzsprung-Russell (HR) diagrams, corresponding to the latter part of the intermediary evolution between the central hydrogen and helium-burning phases. Such stars are characterised by their inert, degenerate cores of helium, surrounded by thick hydrogen-rich envelopes, which support H-burning shells at their bases.

An important observed parameter of the RGB is its maximum (\textit{tip}) luminosity, $L_{\mathrm{RGBT}}$, which is attained at the moment of helium ignition in the star. Its significance comes from the vital role it plays in local determinations of the Hubble parameter and its status as a standard candle of cosmic distance measurement. Consequently, considerable effort has gone into understanding the theoretical processes which underpin it, the canonical picture of which is detailed below. 

After the termination of central hydrogen-burning at the end of the main sequence, the nascent helium core is insufficiently massive in stars below $2M_{\odot}$ to ignite helium. Hydrogen-burning, however, continues in an off-centre shell, the products of which settle onto the core, progressively increasing its mass, temperature and density until the threshold for ignition is met. During this period of inner contraction, the outer envelope initially expands and cools at constant luminosity (the subgiant branch). At a certain point, expansion is no longer physically viable, and the stellar luminosity begins to steadily increase as it ascends the red-giant branch. 

This process is disrupted by the onset of electron degeneracy within the core. In such conditions, energy-loss to neutrino production becomes sizeable in the very centre of the star, causing it to cool. As a result, the peak of the stellar temperature profile shifts outward from the centre, delaying helium ignition and, consequently, allowing $L_{\mathrm{RGBT}}$ to reach higher values than it otherwise would. Neutrino production has been shown to be necessary to reconcile the predictions of stellar models with observed values of $L_{\mathrm{RGBT}}$~\cite{KippenhahnWeigert}.

Through exactly this same mechanism, the presence of any additional energy-loss to, for example, dark photon production, would further delay the onset of helium-burning and lead to larger values of $L_{\mathrm{RGBT}}$. If this energy-loss channel is too efficient, theoretical predictions of $L_{\mathrm{RGBT}}$ will fall outside the observed range, allowing the dark photon parameters to be excluded.

\subsection{RGB-tip constraints on dark photons}
\label{sec: existing RGB constraints}

The two existing RGB-tip dark photon constraints~\cite{Redondo:2013lna, An:2020bxd} have been constructed by demanding that the energy-loss rates per unit mass in Eq.~\ref{eq: energy-loss} be less than $10$~erg g$^{-1}$ s$^{-1}$ when evaluated given average RGB core properties at the moment of helium ignition. We show the limit from Ref.~\cite{An:2020bxd}  in Fig.~\ref{fig: DP parameter space}. The bound derived in Ref.~\cite{Redondo:2013lna} is similar, with some differences due to the using only the L-mode in Ref.~\cite{Redondo:2013lna}, compared with both modes in Ref.~\cite{An:2020bxd}.

This criterion on the energy-loss rate originates in studies of the impact of enhanced energy-loss (assumed to be due to altered neutrino properties) in red-giants~\cite{Sweigert1978,Raffelt:1992pi}. The core mass at helium ignition should not exceed its standard value by more than 5\%, which leads to $\epsilon_{\rm{DP}}\lesssim 2\epsilon_{\nu}$, where $\epsilon_\nu$ is the energy-loss to neutrinos. The results of Ref.~\cite{Sweigert1978} imply that the neutrino luminosity at ignition is approximately $4~\rm{erg}~ g^{-1}~ s^{-1}$, and the standard bound used is $\epsilon_{\rm{DP}}\lesssim 10~\rm{erg} ~g^{-1} s^{-1}$. The energy-loss is calculated for values which represent an average central helium core, namely $\rho=2\times10^5~\rm{g}~cm^{-3}$ and $T=10^8~\rm{K}$. This criterion has been used to bound a variety of models of new physics, including the axion-photon and axion-electron couplings, and neutrino magnetic dipole moments.

We argue that taking an average value for the core parameters will not lead to entirely accurate constraints on dark photons. In particular, consider the resonance condition for transverse dark photons: $\omega_{\rm{pl}}=m_{\rm{DP}}$. Fixing the temperature and density means that this condition will only ever be satisfied for dark photons of a particular mass, corresponding to the peak of the bound in Fig.~\ref{fig: DP parameter space}. However, the core temperature and density evolve upwards throughout the stellar lifetime on the red-giant branch, allowing the resonant production of dark photons with lighter masses. This is not captured by a static model, and leads to stronger constraints on dark photons for masses lower than the resonant peak at $8.6$~keV obtained previously.

On the other hand, for dark photons with larger masses we find that the energy-loss due to dark photons is concentrated so late in the RGB evolution that it does not have time to have an observable impact on the RGB-tip luminosity. The constraints for heavier dark photons are thus less stringent than previously obtained.

In summary, the use of dynamic rather than static models allows our bound to be sensitive to the integrated effects of energy-loss over the entire RGB phase. This, when combined with the existence of resonant production regions unique to dark photons, results in a constraint which is stronger in some regions, but must relax in others.

\subsection{Constructing our constraint}
\label{sec: RGB constraint}

\begin{figure}[t]
    \centering
    \includegraphics[width=0.45\textwidth]{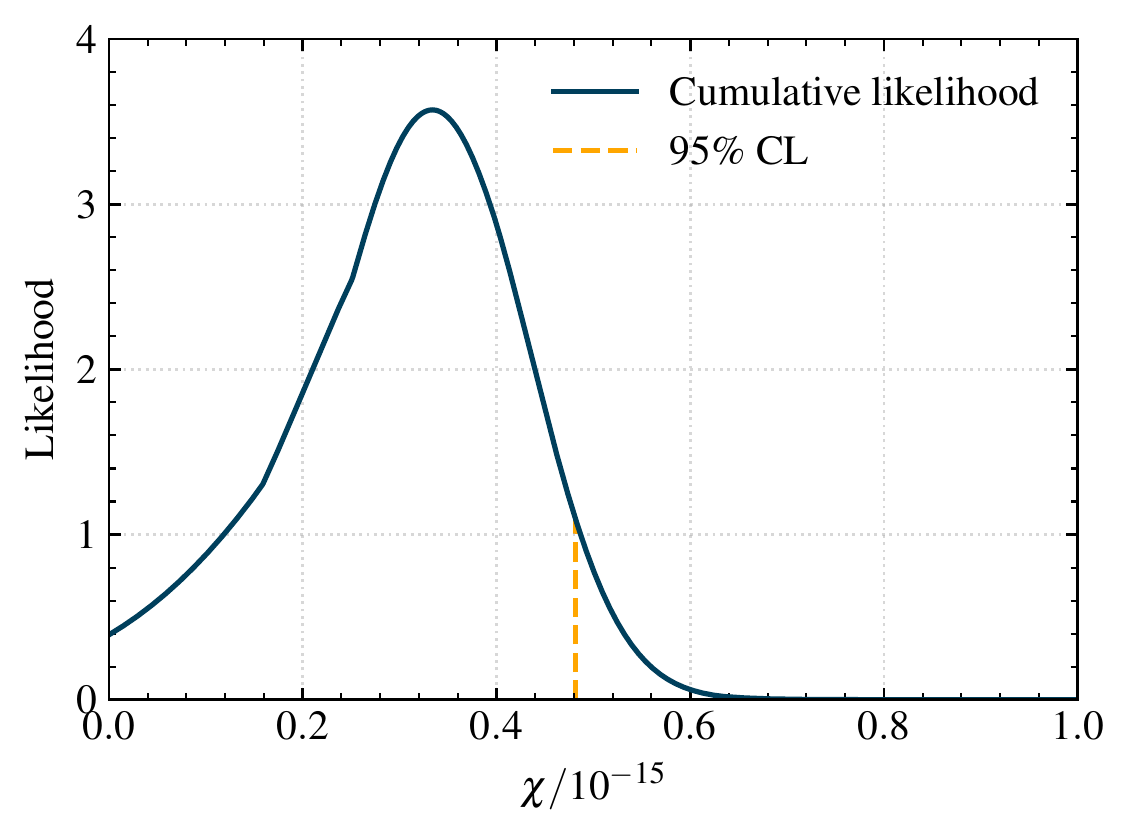}
    \caption{Cumulative likelihood of fit between between observed RGB-tip bolometric magnitudes and those generated by simulation with $m_{\mathrm{DP}}=10$~keV as a function of $\chi$. The 95\% confidence limit is indicated by the dashed yellow line.}
    \label{fig: CLF for 10 keV}
\end{figure}
To quantify the effects of energy-loss to dark photons of a given mass and kinetic mixing strength, we generated predictions for $L_{\mathrm{RGBT}}$ using our edited \texttt{MESA} code for metallicities between $Z=0.0001$ and $Z=0.01$ and compared them to the observed values derived for 22 globular clusters derived in Ref.~\cite{Straniero:2020iyi}. Examples of these predictions, expressed in terms of their \textit{bolometric magnitudes} $V_{\mathrm{bol}}^{\mathrm{RGBT}}$, are shown in Fig. \ref{fig: RGB-T predictions v observations} for a $10$~keV dark photon with $\chi$ ranging between $2.51\times10^{-15}$ and $10^{-15}$, where
\begin{equation}
    \label{eq: bolometric magnitude}
    V_{\mathrm{bol}}=4.74-2.5\log\frac{L}{L_{\odot}}.
\end{equation}
The data from Ref.~\cite{Straniero:2020iyi} are indicated by the black and white stars. The horizontal axis is defined in terms of the global metallicity [M/H], which is related to $Z$ as $[\mathrm{M}/\mathrm{H}]=\log\big(\frac{Z}{X}\big)-\log\big(\frac{Z}{X}\big)_{\odot}$. 

As expected, increasing the kinetic mixing $\chi$ delays helium ignition further, leading to larger RGB-tip luminosities. We note that there there appears to be a preference for non-zero values of $\chi$, but defer comments about this until later in this section.

To determine the quality-of-fit of the theoretical data, we define a cumulative likelihood function for dark photon parameters~\cite{Straniero:2020iyi},
\begin{equation}
    L(m_{\mathrm{DP}}, \chi)=\frac{1}{A}\exp\big( \Delta^{\mathrm{T}} \Sigma^{-1} \Delta \big),
\end{equation}
where $\Delta$ is a vector containing the differences between theoretical and observed luminosities for each of the 22 globular clusters, $A$ is a normalisation constant and $\Sigma$ is the error matrix. The latter includes uncertainty from observation (including statistical and photometric errors), theory as well as chemical composition and cluster age, and accounts for correlation introduced by the theoretical errors, bolometric corrections and the calibration of the distance scale (see Section 3.5 of Ref.~\cite{Straniero:2020iyi}). These errors are included following the method outlined in Ref.~\cite{1994NIMPA.346..306D}. The $1\sigma$ uncertainties for each of the above sources have been selected to match Ref.~\cite{Straniero:2020iyi}

\begin{figure}
    \centering
    \includegraphics[width=0.45\textwidth]{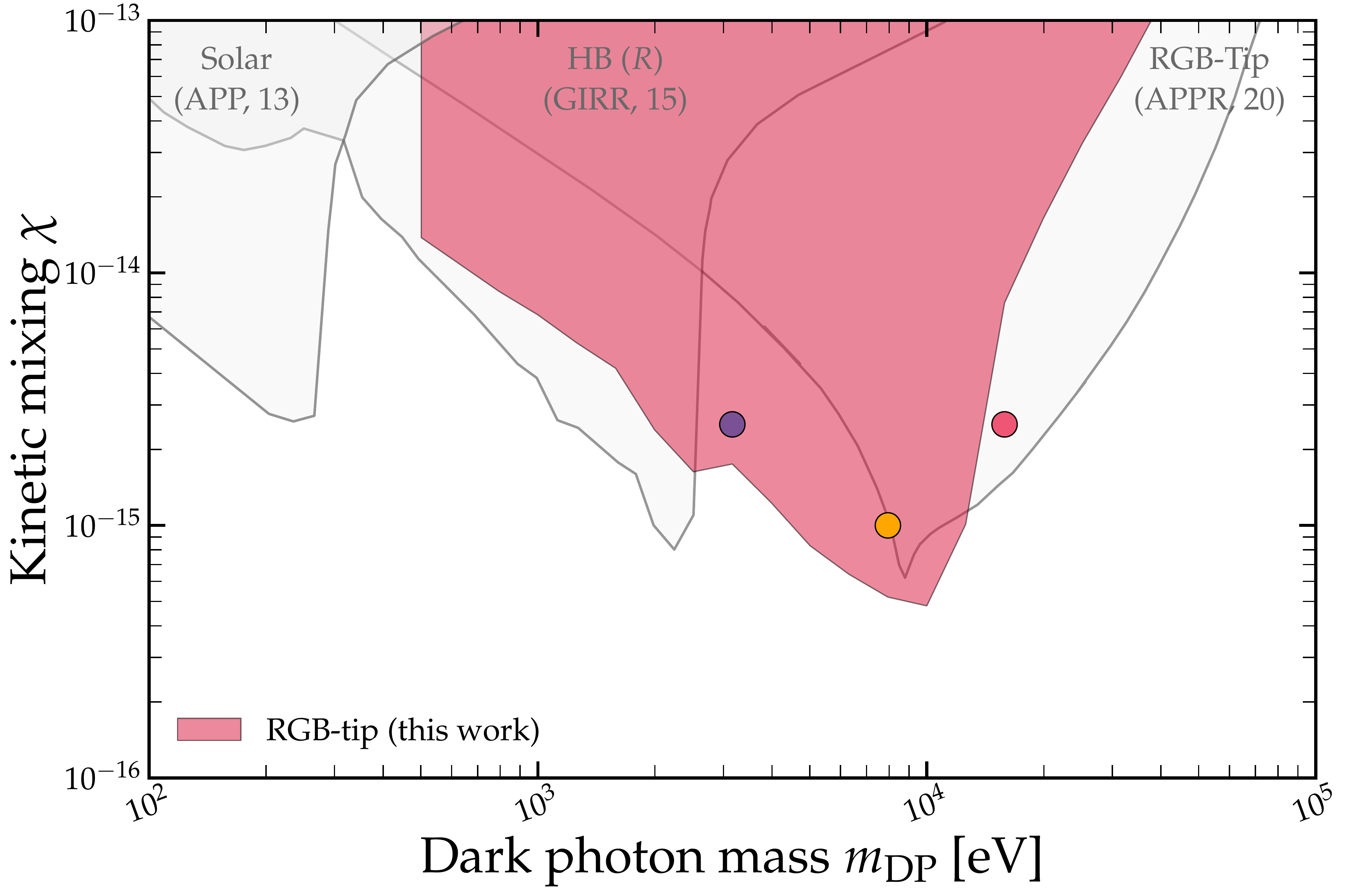}
    \caption{Our new RGB-tip constraint (red) superimposed on top of the solar, $R$-parameter and RGB-tip constraints of Refs.~\cite{An:2013yfc}, \cite{2016JCAP...05..057G} and \cite{An:2020bxd} respectively. See the text for details on discrepancies between these. The three coloured dots correspond to the benchmark points considered in Fig.~\ref{fig: RGB-max-T-evolution} and discussed in the text.}
    \label{fig: RGB-bound}
\end{figure}

An example cumulative likelihood is shown in Fig.~\ref{fig: CLF for 10 keV} for  $m_{\mathrm{DP}}=10$~keV and values of $\chi$ between $0$ and $10^{-15}$. The corresponding 95\% confidence limit on $\chi$ can then be found by solving
\begin{equation}
    \int_{\chi_{95}}^\infty L(m_{\mathrm{DP}}, \chi)d\chi=0.05
\end{equation}
for $\chi_{95}$.
For these data, the implied upper limit is $\chi_{95}=0.48\times10^{-15}$, which is indicated by the dashed orange line  in Fig.~\ref{fig: CLF for 10 keV}.

The preference for non-zero $\chi$ originally noted in Fig.~\ref{fig: RGB-T predictions v observations} is again present in Fig.~\ref{fig: CLF for 10 keV}. While this is not atypical (a preference for a non-zero axion-electron coupling $g_{aee}$ is identified in Ref.~\cite{Straniero:2020iyi}), the effect is considerably stronger in our results. This appears to be due to \texttt{MESA}'s tendency to underestimate $L_{\mathrm{RGBT}}$ compared with both observation and other stellar evolution codes \cite{2022MNRAS.514.3058S}. Consequently, limits of this nature generated using \texttt{MESA} will always be more conservative than those of the latter.

The analysis above was repeated for dark photon masses between $501$~eV and $100$~keV at regular logarithmic intervals. The corresponding excluded region is shown in red in Fig.~\ref{fig: RGB-bound}, superimposed over the existing RGB-tip limit from Ref.~\cite{An:2020bxd}. The lower limit on the scanned masses has been chosen as simulations featuring dark photons lighter than this experience significantly disrupted main sequences, hindering our ability to accurately derive constraints from subsequent evolutionary phases. We shall discuss this more thoroughly in future work. The coloured circles correspond to benchmark scenarios considered in the forthcoming discussion.

While our limit peaks in the same location as Ref.~\cite{An:2020bxd}, there are clear differences between its shape and that of its predecessors. In particular, we exclude a previously unconstrained region between $m_{\mathrm{DP}}\approx500$~eV and $m_{\mathrm{DP}}\approx10$~keV. We also consider that that the limit of Ref.~\cite{An:2020bxd} should relax greatly for $m_{\mathrm{DP}}\gtrsim10$~keV. 

\begin{figure}
    \centering
    \includegraphics[width=0.45\textwidth]{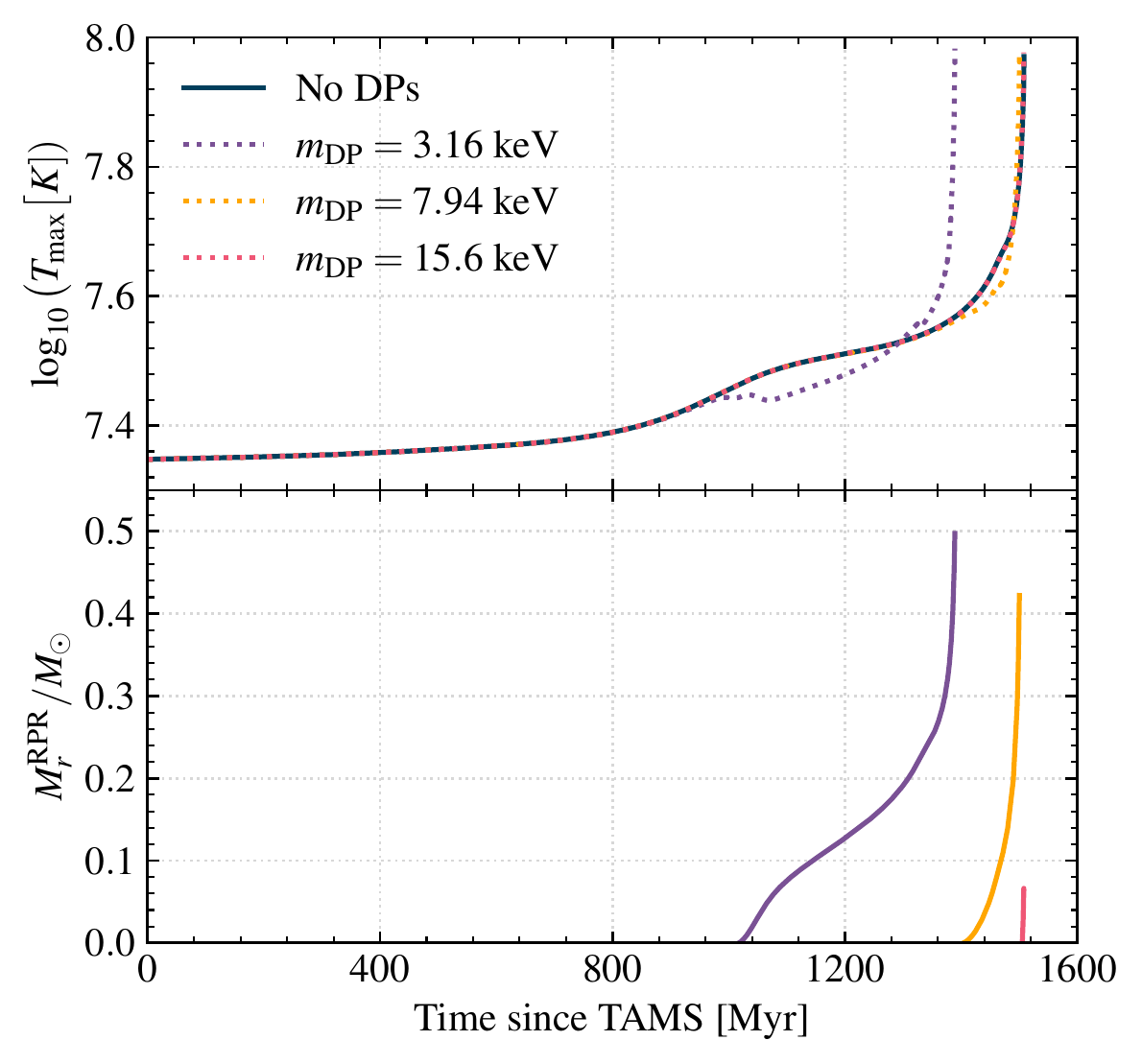}
    \caption{\textit{Top panel:} Evolution of the maximum temperature within a star between the end of the main sequence and the RGB-tip given standard astrophysics alone (dark blue) as well as energy-loss to dark photons with $\chi=10^{-15}$ with $m_{\mathrm{DP}}=3.16$~keV (purple), $7.94$~keV (yellow) and $15.6$~keV (pink). These correspond to the three benchmark points shown in Fig~.\ref{fig: RGB-bound}. \textit{Bottom panel:} RPR radial mass location between the terminal age main sequence and the RGB-tip. Colours match the cases in the top panel. Where no line is visible, an RPR is not yet present in the star.}
    \label{fig: RGB-max-T-evolution}
\end{figure}

To reveal the origin of these discrepancies we consider the evolution of RGB stars associated with the three benchmark points in Fig.~\ref{fig: RGB-bound}. We show this in Fig.~\ref{fig: RGB-max-T-evolution}. In its upper panel, this depicts the evolution of the maximum temperature in an RGB star of initial mass $M_{\mathrm{init}}=0.82$ and $Z=0.001$ for the following four scenarios:
\begin{enumerate}[(i)]
    \item no novel energy-loss (dark blue)
    \item energy-loss to DPs with $m_{\mathrm{DP}}=3.16$~keV and\\ $\chi=2.51\times10^{-15}$ (purple)
    \item energy-loss to DPs with $m_{\mathrm{DP}}=7.94$~keV and\\ $\chi=10^{-15}$ (yellow)
    \item energy-loss to DPs with $m_{\mathrm{DP}}=15.6$~keV and\\ $\chi=2.51\times10^{-15}$ (pink),
\end{enumerate}
while its lower panel provides the radial mass coordinate of the RPR\footnote{i.e. the enclosed mass between $r=0$ and $r=r_{\mathrm{RPR}}$.} as a function of elapsed time since the terminal age main sequence (TAMS) for scenarios (ii), (iii) and (iv). These points are indicated by the purple, yellow and pink circles in Fig.~\ref{fig: RGB-bound}. The purple point is excluded by our limit but not Ref.~\cite{An:2020bxd}, the yellow point excluded by both, and the pink point is excluded by Ref.~\cite{An:2020bxd} but not by us.

In the case of the $7.94$~keV dark photon (yellow), resonant transverse DP emission is not supported in the star until approximately $1400$~Myr after the termination of the main sequence, due to the progressive climbing of its central density/plasma frequency. At this point, considerable energy-loss occurs throughout the core, which causes the evolution of $T_{\mathrm{max}}$ to drop below that of the standard astrophysical simulation, which facilitates the increase in $L_{\mathrm{RGBT}}$ required for the constraint.

Transverse dark photons with $m_{\mathrm{DP}}=3.16$~keV, however, can be resonantly produced much earlier in RGB evolution, approximately $1040$~Myr after the TAMS. It is during the period immediately following this where the majority of the DPs' impact on $T_{\mathrm{max}}$ (and therefore ultimately $L_{\mathrm{RGBT}}$) is experienced. It should come as no surprise then that limits derived from static models of RGB stars at the RGB-tip are insensitive to dark photon effects on this earlier period of post-TAMS evolution.  

Note too that the post-TAMS evolution of the star in case (ii) is shorter by just less than $80$~Myr overall compared with that of the other three cases. This occurs due to the proximity of the RPR to the hydrogen-burning shell for an appreciable length of time after $1200$~Myr, which accelerates nuclear burning activity within the latter. Importantly, this does not offset the helium ignition-delaying effects of this novel energy-loss which are vital for the efficacy of our limit.

Conversely, the simulation including $15.6$~keV dark photons only acquires an RPR very late in its RGB evolution, leaving little time for the effects of this resonant energy-loss to decrease $T_{\mathrm{max}}$ and increase $L_{\mathrm{RGBT}}$, even though it does decrease the central temperature of the star.

 Our results indicate that, despite the presence of these resonant production regions at the RGB-tip for $m_{\rm{DP}}\gtrsim10$~keV, energy-loss from them occurs too late in RGB evolution to affect the key observable $L_{\rm{RGBT}}$. For this reason, we argue that the limit of Ref.~\cite{An:2020bxd} should relax to our new constraint for $m_{\mathrm{DP}}\gtrsim10^4$~keV.

\subsection{Systematic uncertainties}

Given the averaged Gaunt factors $\Bar{g}_{ff}$ vary significantly from the form given in Eq.~\ref{eq: Gbar wo screening} in the degenerate cores of red-giant branch stars, it was necessary to verify that this does not impact the constraint presented in Sec.~\ref{sec: RGB constraint}. A full discussion of this is presented in App.~\ref{app: systematics}.

In investigating this, we confirmed that whenever our limit is driven by resonant production of transverse dark photons, it remains insensitive to the specific value of $\Gamma_{\rm{T}}$ and hence $\Bar{g}_{ff}$. The same is not true for $m_{\rm{DP}}\gtrsim15.8$~keV, where off-resonant production alone is responsible for our constraint. We have adjusted our combined limit in Fig.~\ref{fig: DP parameter space} accordingly in this region of parameter space.

\section{The \texorpdfstring{$R$}{R} and \texorpdfstring{$R_2$}{R2}-parameters}
\label{sec: star cluster counts}
Star cluster counts in globular clusters have a rich history of constraining both stellar astrophysics and Beyond the Standard Model physics. A particularly decorated example is the $R$-parameter,  the ratio of HB to RGB stars in globular clusters, which until recently provided the best bound on the axion-photon coupling for a wide range of masses. More recently the $R$-parameter has been used to constrain dark photons \cite{Redondo:2013lna, An:2013yfc, 2016JCAP...05..057G}.

The success of the $R$-parameter constraint in the context of its original target, axions, is due to the fact that, for the coupling strengths of interest, axions are produced at a greater rate in the helium-burning cores of HB stars than in RGB stars. Any novel energy-loss to these particles therefore shortens the duration of the former, but leaves the latter unaffected, reducing $R$ until it drops below the lower limit from observation. The same is not true for dark photons, however, which may be produced resonantly during both the RGB and HB evolutionary phases. In light of this, recomputing the $R$-parameter constraint using detailed and self-consistent stellar evolution simulations is necessary.

Another limitation on the present dark photon constraints from $R$ is that they do not incorporate the sizeable systematic and stochastic uncertainties introduced by the instability of the horizontal branch convective core boundary. While these issues are widely acknowledged in the astrophysical literature (see e.g. \cite{Lattanzio2}), their impact on axion constraints was not quantitatively studied until a recent paper by the present authors~\cite{Dolan:2022kul}. 

This situation is more delicate for dark photons, owing to possible interplay between the novel energy-loss and the convective structure of the star. Specifically, we find energy-loss to dark photons can spark large and repeated convective episodes during the HB which partially offset the acceleration of the phase or, in some circumstances, actually cause its duration to be extended when compared with the standard astrophysical simulations. Fortunately, another globular cluster parameter, the $R_2$-parameter, the ratio of asymptotic giant branch (AGB) stars to HB stars, is particularly sensitive to these, and can be used to set a strong limit on their existence and hence on dark photons.

The goal of this section is therefore to construct new bounds on dark photons from both $R$ and $R_2$. First, however, we shall discuss the specific effects which energy-loss to dark photons has on the HB phase, owing to the latter's relevance for both bounds.

\subsection{Dark photon effects on horizontal branch stars}

\subsubsection{Standard HB evolution}
\label{sec: The HB}
\begin{figure*}[t]
    \centering
    \includegraphics[width = 0.9\textwidth]{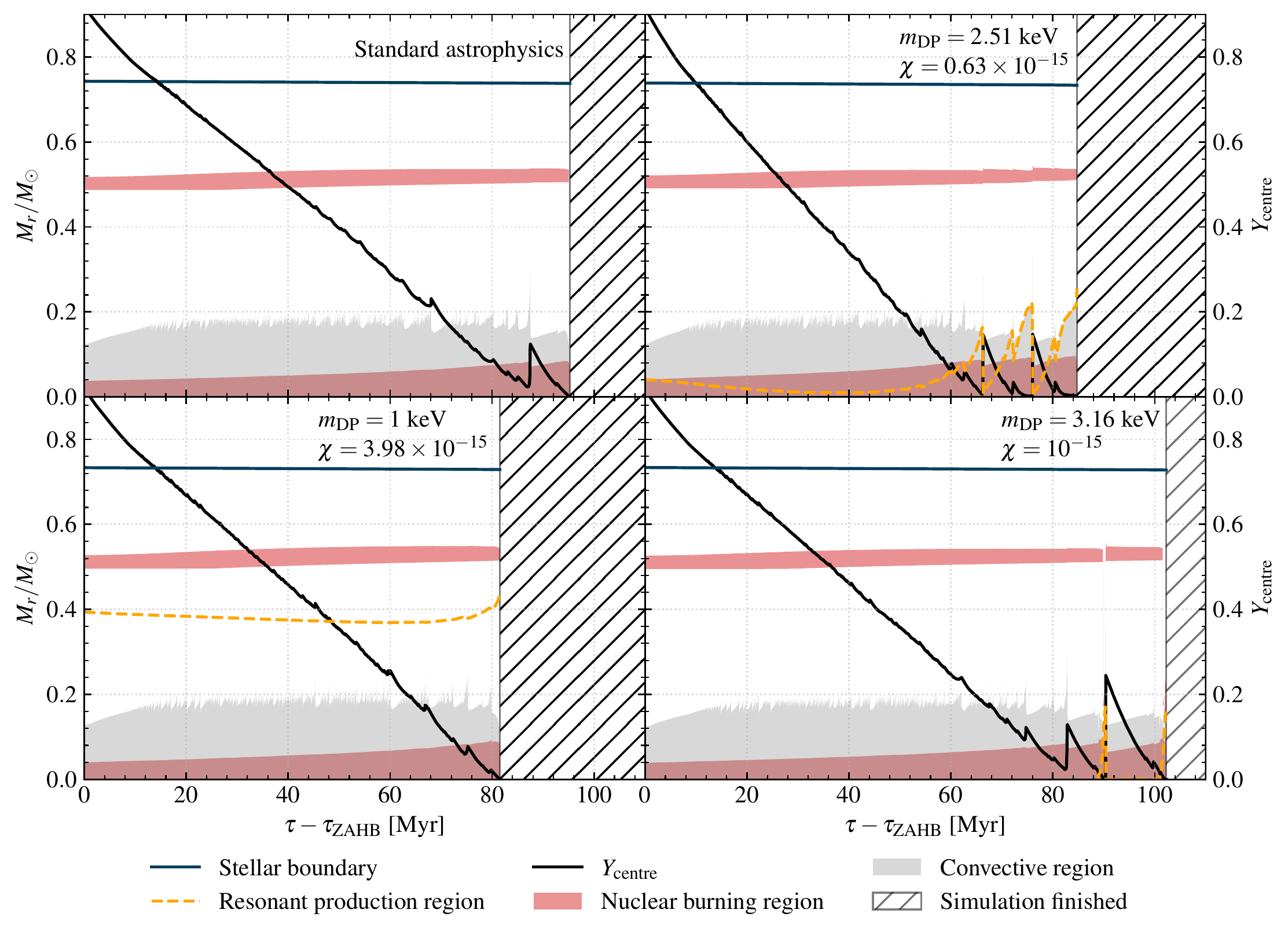}
    \caption{Kippenhahn diagrams showing the evolution along the horizontal branch for (counterclockwise from top-left) standard astrophysics, and with dark photons of mass 1, 3.16 and 2.51~keV. The lower (upper) red regions show helium (hydrogen) burning regions, and the grey regions show the convective core. The black lines show the central helium mass fraction $Y_{\rm{centre}}$. The orange lines show the location of the resonant production regions. Finally, the outer boundary of the star is the dark blue horizontal line at $M_r/M_{\odot}\approx 0.75$.}
    \label{fig: HB Kippenhahns}
\end{figure*}

The horizontal branch is the evolutionary phase which immediately follows the onset of helium burning at the RGB-tip in low mass stars ($\lesssim2M_{\odot}$). At the centre of these stars is a convective helium-burning core, embedded within a larger helium-rich zone left over from the previous hydrogen-burning phases. This, in turn, is surrounded by a thick hydrogen-rich envelope at the base of which is a thin shell supporting hydrogen burning.

 An example of the structural evolution of an HB star with $M_{\mathrm{init}}=0.82M_{\odot}$ is shown in the Kippenhahn diagram in the top left panel of Fig.~\ref{fig: HB Kippenhahns}. Here, the inner (outer) red areas indicate stellar regions of strong helium (hydrogen) burning, while the grey area denotes the convective core. The outer boundary of the star is indicated by the dark blue horizontal line at $M_r/M_{\odot}\approx 0.75$.

During the HB, nuclear burning gradually replaces the helium in the core with carbon and oxygen until the helium supply is exhausted. In light of this, the central mass fraction of helium, $Y_{\mathrm{centre}}$, is also included in Fig.~\ref{fig: HB Kippenhahns} (black line) to indicate the progression of the phase. We end the simulation when the central helium mass-fraction reaches zero, indicated by the hatched region on the right-hand side.

A well-known consequence of this chemical evolution is the instability of the core's convective boundary, which manifests in repeated episodes of growth and splitting (see App.~\ref{app: HB CBPs} for further details). Such episodes can be seen as the spiky protuberances above the convective region in Fig.~\ref{fig: HB Kippenhahns}. These also explain why the overall decrease of $Y_{\mathrm{centre}}$ throughout the phase is punctuated by regular periods of increase as the core forays into the helium-rich region.

The most dramatic impact of this instability occurs when additional helium is deposited into the core near the end of the HB phase. In this scenario, the introduction of new helium into a core which is starved of nuclear fuel results in a short burst of strong nuclear burning. This, in turn, seeds a large convective excursion into the helium-rich zone, which floods the core with fresh helium, elongating the phase. Such events are known as \textit{core breathing pulses} (CBPs). An example of a CBP is visible in the top left panel of Fig.~\ref{fig: HB Kippenhahns} approximately $88$~Myr after the zero-age horizontal branch (ZAHB).

The number and duration of CBPs predicted in HB simulations is not computationally stable and the example shown in the top left panel of Fig.~\ref{fig: HB Kippenhahns} has been selected for illustrative purposes. For this reason, when deriving bounds from this phase in the subsequent sections, we repeat HB simulations with different spatial and temporal resolutions to sample the resulting stochastic variation in phase duration. We note that, while it is still debated whether CBPs are physical or merely numerical artifacts, a conservative analysis should account for their possible existence. 

\subsubsection{Dark photons and the HB}
\label{sec: DPs and the HB}

\begin{figure}[t]
    \centering
    \includegraphics[width=0.45\textwidth]{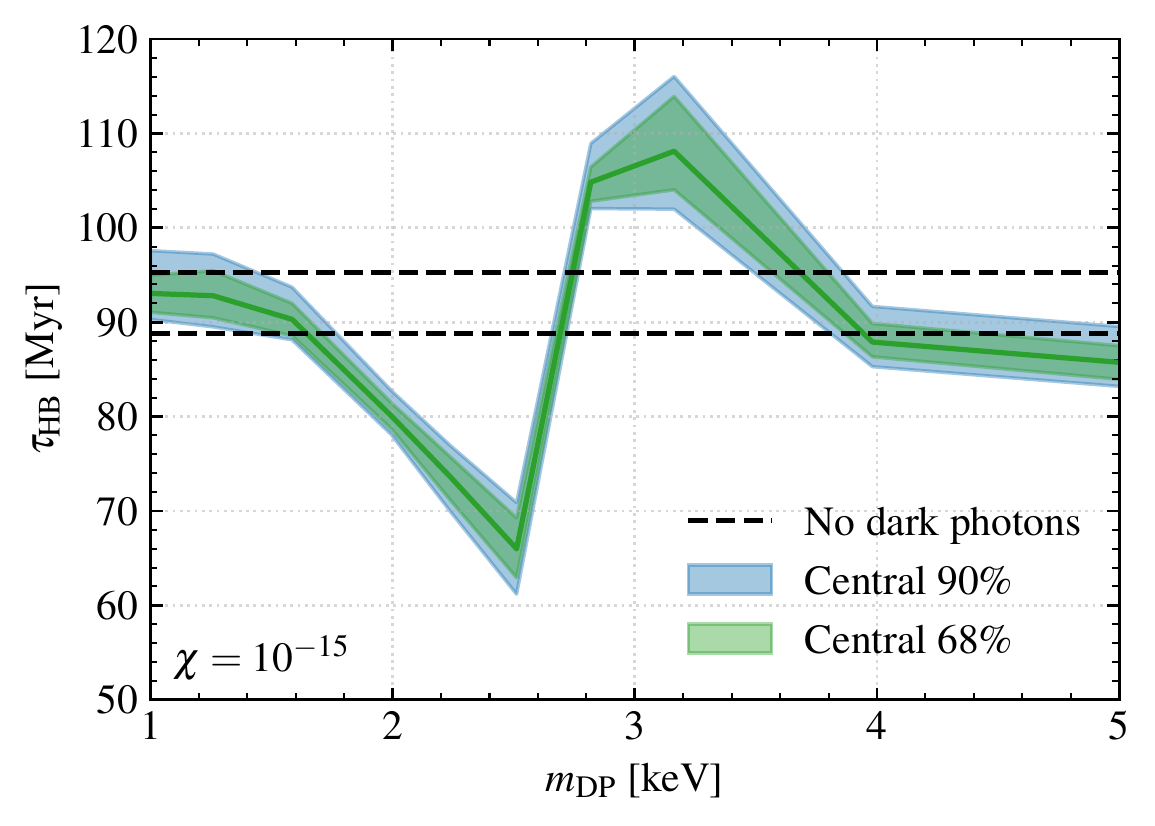}
    \caption{This figure shows the impact of dark photon energy-loss on the horizontal branch lifetime $\tau_{\mathrm{HB}}$ when $m_{\rm{DP}}$ is varied. We have fixed the mixing parameter $\chi=1\times10^{-15}$ and stellar mass to $0.82M_{\odot}$, and run 100 simulations for each mass. The blue and green bands show the central 90\% and 68\% of runs respectively. The horizontal dashed lines show the range of $\tau_{\mathrm{HB}}$ for standard astrophysics.}
    \label{fig: HB lifetimes}
\end{figure}

The impact of transverse dark photon production on the HB depends on the proximity of the resonant production region to the convective core throughout the phase.

We shall discuss these differences by comparing the predictions of simulations given three different choices of the dark photon parameters:
\begin{enumerate}[(i)]
    \item $m_{\mathrm{DP}}=1$~keV and $\chi=3.98\times10^{-15}$
    \item $m_{\mathrm{DP}}=2.51$~keV and $\chi=0.63\times10^{-15}$
    \item $m_{\mathrm{DP}}=3.16$~keV and $\chi=10^{-15}$,
\end{enumerate}
which have been selected to illustrate qualitatively different effects of dark photon energy-loss on the HB.

Examples of HB evolution given for each of these cases are shown in the remaining three panels of Fig.~\ref{fig: HB Kippenhahns}. We show the location of the resonant production region in each diagram as a dashed orange line. 

In case (i), the RPR stays in the vicinity of $M_r=0.4M_{\odot}$ for the duration of the HB, far from the convective core. We observe that the presence of novel energy-loss has accelerated HB evolution, even though the RPR is not located within the helium burning core. Aside from this, and the absence of a CBP in this simulation, the evolution qualitatively resembles that given standard astrophysics alone.

As we increase $\chi$, resonant energy-loss is substantial enough to disrupt the standard temperature gradient in the resonant production region, causing a small convective region to form around it. As $\chi$ is increased further, the size of this region increases until it ultimately merges with the inner convective core. This results in an enlarged convective core, the boundary of which is set by the location of the RPR. While non-standard, such stars do not suffer from the convective instabilities which normally plague the HB.

In case (ii) the RPR is within the burning core for the vast majority of the HB, which initially accelerates the progression of the phase, as expected. As the supply of helium in the core dwindles, the star's density profile -- and plasma frequency profile -- shifts upwards, causing the RPR to drift further from the centre. This outward movement of the RPR through the convective core seeds large convective episodes at $\tau-\tau_{\mathrm{ZAHB}}\approx66$~Myr and $76$~Myr, which flood the core with fresh helium. The net effect of these is the elongation of the HB from approximately $66$~Myr to $82$~Myr for the example in Fig.~\ref{fig: HB Kippenhahns}.

Though these events have a similar effect on HB evolution to the aforementioned core breathing pulses, these two scenarios should be distinguished from one another as, unlike the latter, the former are sparked by the movement of the RPR rather than the chemical evolution of the core. We therefore adopt the name \textit{dark photon core breathing pulses} (DP-CBPs) for these and detail their exact formation mechanism, contrasting it to that of standard astrophysical CBPs, in App.~\ref{app: HB CBPs}.

Finally, in case (iii), no RPR is present in the star until near the terminal age horizontal branch (TAHB), meaning no appreciable acceleration of this phase occurs. Nevertheless, once the star gains an RPR, its outward movement again seeds a DP-CBP, which results in the overall elongation of the HB phase compared with that given standard astrophysics alone.

\subsubsection{Impact on HB duration}

\begin{figure*}
    \centering
    \includegraphics[width=0.95\textwidth]{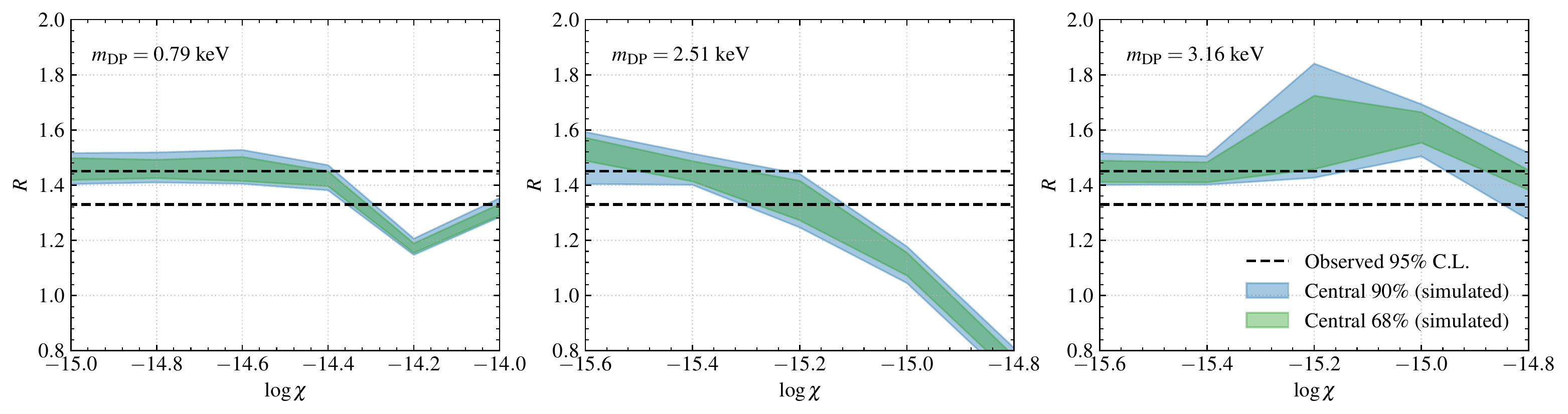}
    \caption{From left to right, the variation of the $R$-parameter for $m_{\mathrm{DP}}$=0.79~keV, 2.51~keV and 3.16~keV for varying $\chi$. We have run 100 simulations for each mass. The blue and green bands show the envelope of the central 90 and 68 runs respectively. The horizontal dashed lines show the observed 95\% confidence interval of $R$. Note the different x-axis ranges in each sub-panel.}
    \label{fig: R examples}
\end{figure*}

As predictions of the HB duration are not computationally stable, it is important to examine the effects of dark photon energy-loss on the full range of possible predictions for $\tau_{\mathrm{HB}}$. To do this, we simulated the HB evolution of $0.82M_{\odot}$ stars including energy-loss to dark photons with $\chi=10^{-15}$ and $m_{\mathrm{DP}}$ ranging from $1$~keV to $5.01$~keV 100 times each, employing different spatial and temporal resolutions in each run. The central 90\% and 68\% of generated outcomes are shown as the blue and green regions in Fig.~\ref{fig: HB lifetimes} respectively. The average is given by the solid green line and the upper and lower ranges of $\tau_{\mathrm{HB}}$ given no novel energy-loss are indicated by the dashed black lines. 

Clearly, for $1$~keV dark photons, a value of $\chi=10^{-15}$ is insufficient to appreciably accelerate the phase, and the resulting HB lifetimes agree well with those from standard astrophysics alone. Increasing $m_{\mathrm{DP}}$ from this point, while keeping $\chi$ constant, has two ramifications: (i) the magnitude of energy-loss increases (c.f. Eqs.~\ref{eq: L production} and \ref{eq: T production}), and (ii) the RPR moves towards the centre of the star.

Initially, only the former has a meaningful impact, and the predicted $\tau_{\mathrm{HB}}$ values fall and cluster more tightly. However, as $m_{\mathrm{DP}}$ approaches $2.24$~keV, we begin to see interplay between the energy-loss and convective structure (i.e. DP-CBPs occur), which manifests as an increase in the thickness of the shaded region(s). 

At larger masses still, no resonant production region is present during the majority of the HB phase, meaning that little acceleration of the phase occurs. Despite this, the outward movement of the resonant production region, once it enters the star near the TAHB, may still spark dark photon core breathing pulses as in the bottom right panel of Fig.~\ref{fig: HB Kippenhahns}. This results in predictions for $\tau_{\mathrm{HB}}$ which are substantially larger than those given astrophysics alone, even with standard CBPs present. Finally, for $m_{\mathrm{DP}}\gtrsim4$~keV, HB stars no longer acquire RPRs early enough for any DP-CBPs to be present in the simulation, and the predictions again resemble those given standard astrophysics, albeit with a marginal offset caused by off-resonant DP production in the HB core.

With the these trends in mind, we now present our bounds from $R$ and $R_2$.

\subsection{The \texorpdfstring{$R$}{R}-parameter}
\label{sec: R-parameter}

Observationally, the $R$-parameter is defined as
\begin{equation}
    R=\frac{N_{\mathrm{HB}}}{N_{\mathrm{RGB}}},
\end{equation}
where $N_{\mathrm{HB}}$ is the number of horizontal branch stars, and $N_{\mathrm{RGB}}$ is the number of red-giant branch stars brighter than the F555W filter magnitude at the ZAHB, $V_{\mathrm{ZAHB}}$, which is taken to be at $\log T_{\mathrm{eff}}=3.85$.

\subsubsection{Calculating \texorpdfstring{$R$}{R}}
\label{sec: calculating R}

If we assume that both the red-giant and horizontal branches are well-populated, we may approximate $R$ theoretically as:
\begin{equation}
    R=\frac{\tau_{\mathrm{HB}}}{\tau_{\mathrm{RGB}}},
\end{equation}
where $\tau_{\mathrm{HB}}$ ($\tau_{\mathrm{RGB}}$) is the duration of the HB (RGB) for a single stellar track. The HB lifetime is defined as the elapsed time between the ZAHB and TAHB and is extracted from our stellar evolution simulations.

In order for our theoretical values of $R$ to adequately represent the observed equivalent, we define $\tau_{\mathrm{RGB}}$ as the elapsed time on the RGB between the point at which $V=V_{\mathrm{ZAHB}}$ and the RGB-tip. Finding the former point requires the conversion of our theoretical HB and RGB stellar tracks to F555W magnitudes via the relation:
\begin{equation}
    V(L, T_{\mathrm{eff}}, \log g)=V_{\mathrm{bol}}(L)-V_{\mathrm{BC}}(T_{\mathrm{eff}}, \log g),
\end{equation}
where the bolometric magnitude $V_{\mathrm{bol}}$ was given in Eq.~\ref{eq: bolometric magnitude} and $V_{\mathrm{BC}}$ are a set of appropriate bolometric corrections (BCs). In this work, we have employed the HST Wide Field and Planetary Camera 2 F555W BC tables available at the \texttt{MESA} Isochrones and Stellar Tracks (MIST) website\footnote{\url{https://waps.cfa.harvard.edu/MIST/model\_grids.html}}, which were computed from the CK3 grid.

Furthermore, our calculations must also reflect the fact that $V_{\mathrm{ZAHB}}$ is defined at $\log T_{\mathrm{eff}}=3.85$. A complicating factor here is the treatment of mass loss on the RGB. In globular cluster HR diagrams, horizontal branches contain a certain degree of thickness, which can partially be attributed to scatter in the rate of mass-loss due to strong stellar winds in the late-RGB. 

Theoretically, we can accommodate this by performing our simulations for several different value of the Reimers parameter $\eta_{\mathrm{R}}$, which sets the scale of mass-loss during the RGB. For increasing $\eta_{\mathrm{R}}$, the HB hydrogen-envelope becomes smaller, shifting $\log T_{\mathrm{eff}}^{\mathrm{ZAHB}}$ upwards. For appropriately chosen values of $\eta_{\mathrm{R}}$, our predicted values of $\log T_{\mathrm{eff}}^{\mathrm{ZAHB}}$ can straddle the desired value of $3.85$ and we can interpolate between our results to obtain $R$ in a manner which reflects the observed value. This is similar to the approach of Ref.~\cite{Ayala:2014pea}, except we sample $\eta_{\mathrm{R}}$ more finely.

\subsubsection{Results}
\label{sec: R Results subsec}

In Fig.~\ref{fig: R examples}, we show examples of the variation of $R$ with $\chi$ for three different dark photon masses: $m_{\mathrm{DP}}=0.79$~keV, $2.51$~keV and $3.16$~keV. The non-zero width of these regions originates from the  stochastic variation of $\tau_{\mathrm{HB}}$ (due to the presence of core breathing pulses), which we have sampled by repeating each HB simulation 100 times for different spatial and temporal resolutions. The blue and green regions correspond to the central 90\% and 68\% of predicted values for $R$, while the black dashed lines indicated the observed 95\% confidence limits from Ref.~\cite{Ayala:2014pea}. The input physics adopted in these simulations is given in Appendix D.1 of our previous paper \cite{Dolan:2022kul}.

The simplest scenario is reflected in the middle panel, where $m_{\mathrm{DP}}=2.51$~keV, which places the resonant production region in the core for the entire duration of the HB phase. In this scenario increasing $\chi$ while leading to some DP CBPs, nevertheless gradually reduces $\tau_{\mathrm{HB}}$, while $\tau_{\mathrm{RGB}}$ is only minimally affected, leading to a reduction in $R$. The inner 90\% of data falls below the observed 95\% CL at $\chi=0.76\times10^{-15}$, at which point we define our constraint. 

When $m_{\mathrm{DP}}=0.79$~keV (left panel), although the RPR is not within the core we nevertheless see a progressive decrease in $R$ from the accelerated consumption of helium in the HB core which becomes inconsistent with observed limits at $\chi=5.4\times10^{-15}$. However, if $\chi$ is increased further from this point, resonant dark photon energy-loss becomes substantial enough to affect stellar evolution during the RGB (c.f. Fig.~\ref{fig: RGB-bound}).

Specifically, the increase in helium-core mass at the point of helium-ignition leads to larger ZAHB luminosities, and hence lower values of $V_{\mathrm{bol}}^{\mathrm{ZAHB}}$. This, combined with accelerated hydrogen-burning timescales caused by the presence of the resonant production region immediately below the H-burning shell, leads to a reduction in $\tau_{\mathrm{RGB}}$ and increase in $R$. By $\chi=8.51\times10^{-15}$, our predictions for $R$ are back within the range indicated by observation and our constraint relaxes. The value of $\chi$ for which our bound relaxes overlaps with our RGB-tip constraint, leaving our overall limit unaffected.

Finally, for $m_{\mathrm{DP}}=3.16$~keV, dark photon core breathing pulses late in the HB, which begin to occur for $\chi>0.39\times10^{-15}$, give rise to a large range in predictions for $\tau_{\mathrm{HB}}$ and hence $R$. This, combined with the absence of an RPR in the HB for the majority of the phase, conspire to increase predictions of $R$ such that it never drops below the observed range and no limit can be defined.

\subsubsection{The \texorpdfstring{$R$}{R}-constraint}
\label{sec: R Results}
\begin{figure}
    \centering
    \includegraphics[width=0.47\textwidth]{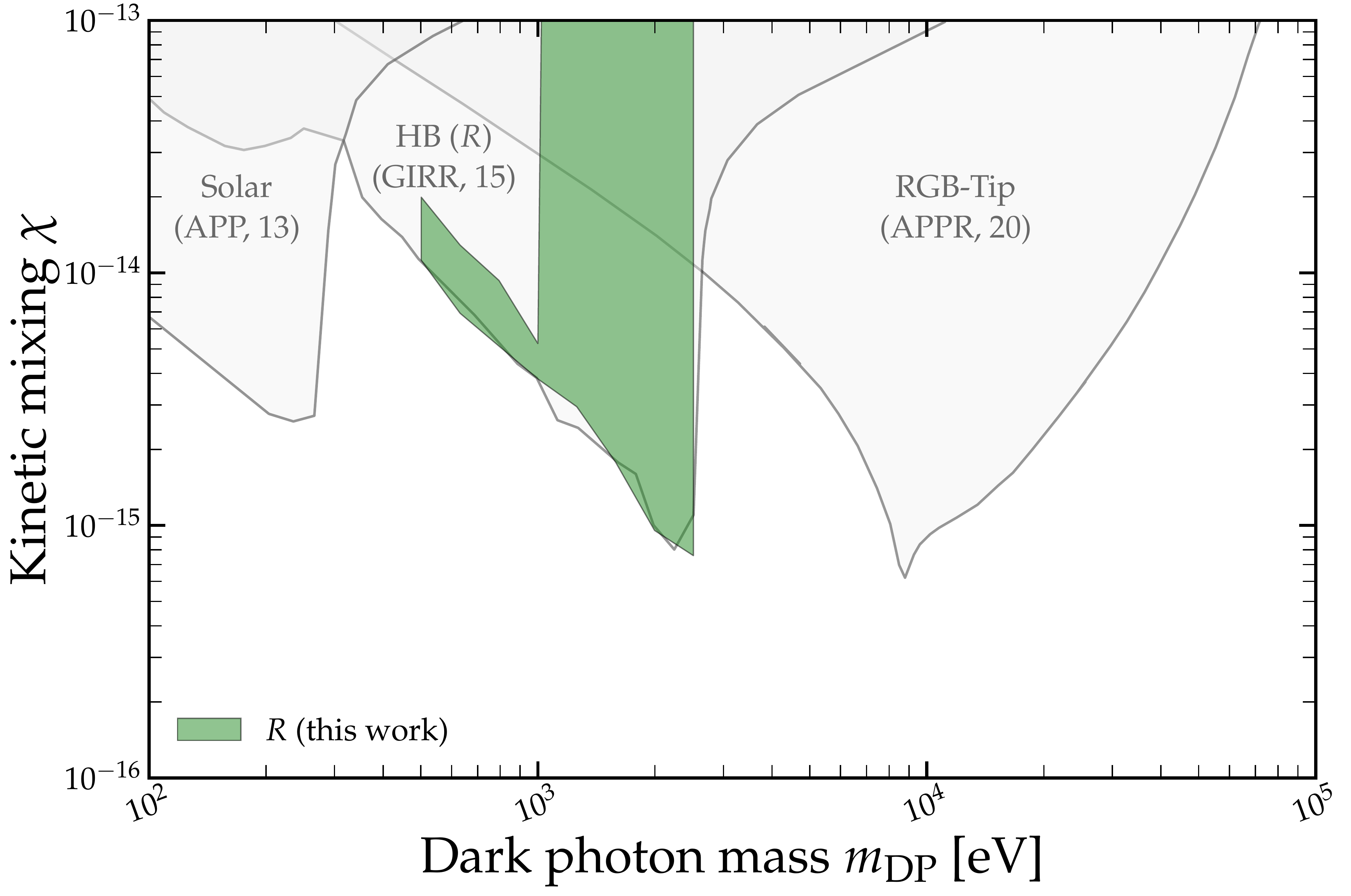}
    \caption{Our new $R$ constraint (green) superimposed on top of the solar, $R$-parameter and RGB-tip constraints of Refs.~\cite{An:2013yfc}, \cite{2016JCAP...05..057G} and \cite{An:2020bxd} respectively.}
    \label{fig: R bound}
\end{figure}

We construct our constraint, which is shown in green in Fig.~\ref{fig: R bound}, by scanning at logarithmic intervals in dark photon mass between $m_{\mathrm{DP}}=501$~eV and $3.16$~keV. This lower limit has again been chosen due to complications in simulations from disrupted main sequence evolution, while dark photons more massive than the upper limit are not produced resonantly during the HB.

Our obtained limit generally agrees well with the existing bound from Ref.~\cite{2016JCAP...05..057G}. The obvious exception is when $m_{\mathrm{DP}}\lesssim1$~keV, at which point dark photon emission during the RGB phase causes the constraint to relax at larger values of $\chi$ as in the left-panel of Fig.~\ref{fig: R examples}. For the datapoints between $m_{\rm{DP}}=501$~eV and $794$~eV, this relaxation occurs at values of $\chi$ already ruled out by our RGB-tip constraint. 

Dark photons with $m_{\rm{DP}}=1$~keV, however, are uniquely positioned to be able to accelerate RGB evolution \textit{and} disrupt standard convective structure during the HB. These effects on different evolutionary phases conspire to reduce the efficacy of the $R$-parameter limit, causing it to relax below the RGB-tip (and $R_2$) constraints. This gives rise to the small unconstrained triangle in our total limit, visible in Fig.~\ref{fig: DP parameter space}.

\subsection{The \texorpdfstring{$R_2$}{R2}-parameter}
\label{sec: R2-parameter}
\begin{figure*}
    \centering
    \includegraphics[width=0.95\textwidth]{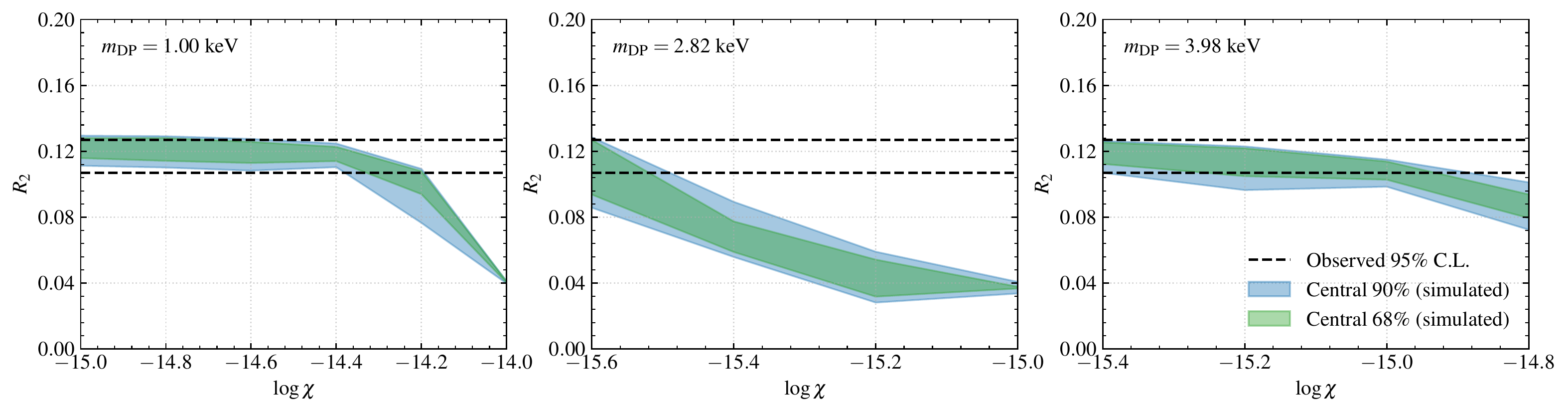}
    \caption{From left to right, the variation of the $R_2$-parameter for $m_{\mathrm{DP}}$=1.00~keV, 2.82~keV and 3.98~keV for varying $\chi$. We have run 100 simulations for each mass. The blue and green bands show the envelope of the central 90 and 68 runs respectively. The horizontal dashed lines show the observed 95\% confidence interval of $R_2$. Note the different x-axis ranges in each sub-panel.}
    \label{fig: R2 distributions}
\end{figure*}

As previously mentioned, the $R_2$-parameter is the ratio of asymptotic giant branch to horizontal branch stars in globular clusters.

\subsubsection{Observed values of \texorpdfstring{$R_2$}{R2}}
The AGB is the evolutionary phase immediately following the termination of central helium-burning on the HB. At the centre of an AGB star is an inert (non-burning) core composed of carbon and oxygen. This sits beneath successive helium and hydrogen-rich envelopes, which support helium- and hydrogen-burning at their respective bases. 

In the initial sub-phase of the AGB, termed the early AGB or E-AGB, the He-B shell moves outward through the He-rich zone, steadily producing more carbon and oxygen, until the entire supply of helium in this layer has been exhausted. This then gives way to successive periods of quiescent hydrogen-shell burning, which re-establishes some supply of helium, punctuated by thermally unstable helium-shell flashes. This second sub-phase, termed the thermally pulsating (TP)-AGB, continues until stellar winds have stripped away the outer layers of the star, leaving a proto-white dwarf in its wake.

The observed values of $R_2$ reported in Ref.~\cite{Lattanzio2} were determined by first converting globular cluster HR diagrams to probability density functions of $\Delta\log L=\log L-\log L_{\mathrm{HB}}$, where $\log L_{\mathrm{HB}}$ is the mode of the HB luminosity distribution (see Fig.~6 of Ref.~\cite{Lattanzio2}). These distributions have two peaks -- the first corresponding to the HB and the second to the AGB -- which are separated by a clear minimum at some log luminosity difference $\Delta\log L_{\mathrm{min}}$. 

Up to scaling by a normalisation constant, $N_{\mathrm{AGB}}$ is then given as the area underneath the curve between $\Delta\log L=\Delta\log L_{\mathrm{min}}$ and $\Delta\log L=1.0$, while $N_{\rm{HB}}$ is given by the area to the left-hand side of $\Delta\log L_{\mathrm{min}}$. $R_2$ is the ratio between these two areas. Using this method, the authors of Ref.~\cite{Lattanzio2} determined $R_2$ to be $0.117\pm0.005$ ($1\sigma$).

Given only AGB stars with $\Delta\log L\leq1.0$ contribute to this definition of $R_2$, the E-AGB phase alone will be relevant for our theoretical calculations. Consequently, any acceleration in the evolutionary progression of the E-AGB, which is governed by the activity of the He-B shell, will decrease $R_2$ until it no longer agrees with the observed limit at which point a constraint can be defined. As we shall see, the efficiency with which this occurs is subject to changes in the HB duration as well.

\subsubsection{Calculating \texorpdfstring{$R_2$}{R2} theoretically}
\label{sec: R2 calculations}

Our methods for calculating $R_2$ from stellar simulations, following that of Ref.~\cite{Lattanzio2}, were detailed thoroughly in our recent paper \cite{Dolan:2022kul}. In light of this, we will only repeat the key points here, i.e.\ that $R_2$ can be extracted from a single stellar track by
\begin{enumerate}
    \item Converting the HB and AGB HR diagrams to the probability density function of $\Delta \log L$ using Eq. 7 of Ref.~\cite{Lattanzio2},
    \item Finding the value of $\Delta\log L_{\mathrm{min}}$, i.e. the log luminosity difference of the minimum between the HB and AGB peaks in the PDF,
    \item Computing $R_2$ as the ratio of the area to the right and left of $\Delta\log L_{\mathrm{min}}$.
\end{enumerate}

In Fig.~\ref{fig: R2 distributions} we show the variation of $R_2$ as a function of $\chi$ for $1$~keV, $2.82$~keV and $3.98$~keV dark photons. The non-zero width of these regions again originates from the stochastic variation of $\tau_{\mathrm{HB}}$ due to both DP and astrophysical CBPs and has been generated by repeating the HB and AGB simulations 100 times each with different spatial and temporal resolutions. The blue and green regions again correspond to the central 90\% and 68\% of simulations respectively. The input physics adopted in these simulations is the same as that of Sec.~\ref{sec: R-parameter} and was discussed comprehensively in Ref.~\cite{Dolan:2022kul}.

In the first panel $m_{\mathrm{DP}}=1$~keV and resonant production is supported during both the HB and AGB phases. Consequently, any decrease in the duration of the former is compensated by a similar reduction in the latter leaving $R_2$ invariant until $\log{\chi}=-14.4$. As $\chi$ is increased further a second convective region begins to form during the HB in the region around the RPR, which merges with the convective core and the simulations converge. Such stars have enlarged He-depleted cores at the beginning of the E-AGB, which shortens $\tau_{\mathrm{AGB}}$ and causes $R_2$ to fall below the allowed range.

When $m_{\mathrm{DP}}=2.82$~keV, the presence of dark photon core breathing pulses in the simulations causes a large spread in the duration of $\tau_{\mathrm{HB}}$. Interestingly, stars which undergo more CBPs have larger helium-depleted cores at the beginning of the E-AGB, which shortens its duration even before the additional effects of novel energy-loss are taken into account. This compounding effect results in a very wide spread in predictions for $R_2$ and a strong bound of $\chi=3.24\times10^{-16}$ when the inner 90\% of data fall below the observed limit.

Finally, for $m_{\mathrm{DP}}=3.98$~keV, no resonant production region is supported within the HB core for the entire duration of that evolutionary phase and no dark photon core breathing pulses appear in our simulations. In these conditions, $R_2$ acts purely as a probe of the progression on the E-AGB, which can be accelerated due to the continued presence of resonant dark photon energy-loss. It is this which causes the $R_2$ distributions to slowly decrease, maintaining an approximately constant width, until they become inconsistent with observation at $\chi=1.31\times10^{-15}$.

\subsubsection{The \texorpdfstring{$R_2$}{R2}-parameter constraint}
\begin{figure}
    \centering
    \includegraphics[width=0.45\textwidth]{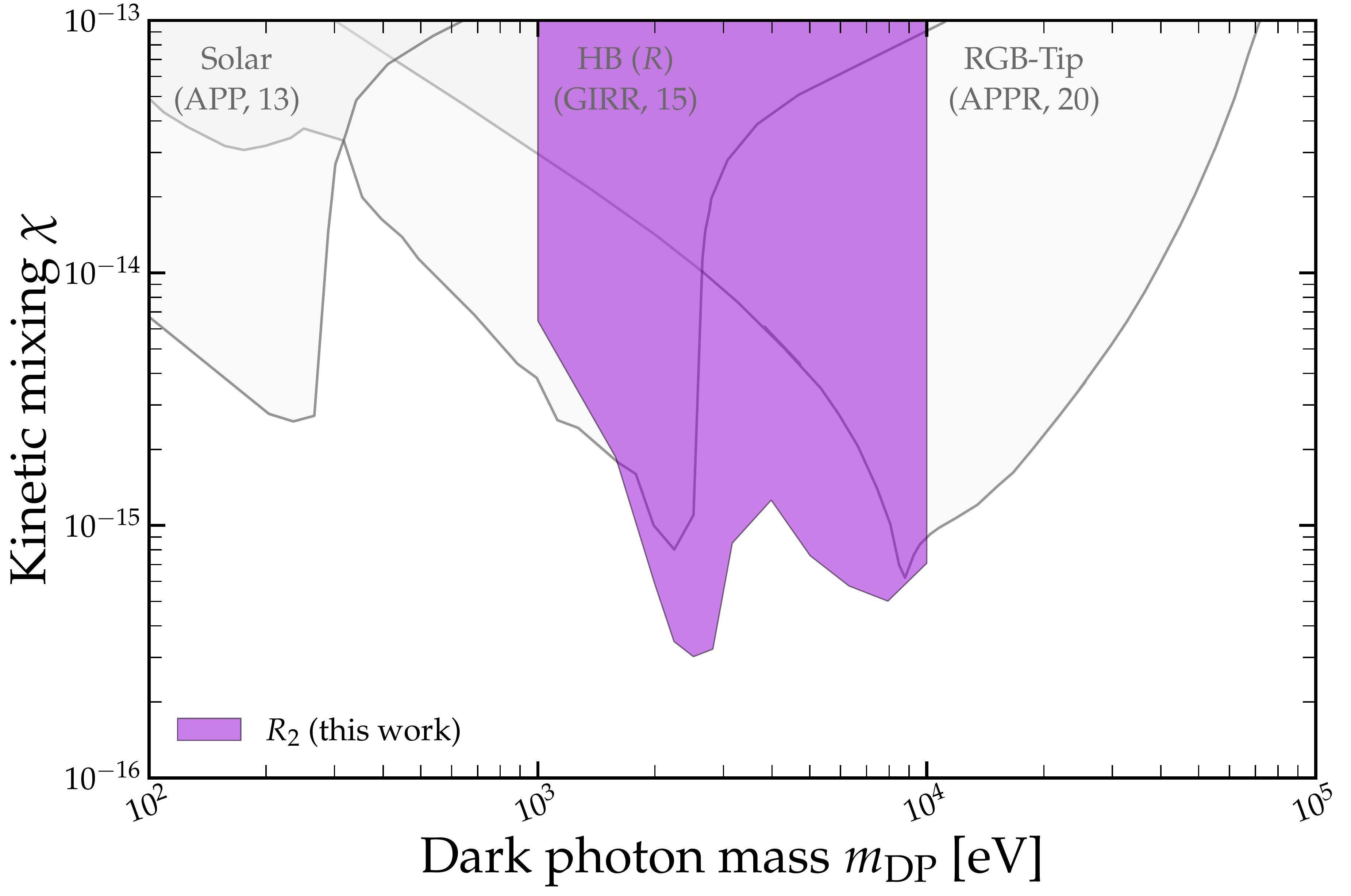}
    \caption{Our new $R_2$ constraint (purple) superimposed on top of the solar, $R$-parameter and RGB-tip constraints of Refs.~\cite{An:2013yfc}, \cite{2016JCAP...05..057G} and \cite{An:2020bxd} respectively.}
    \label{fig: R2 bound}
\end{figure}

To generate our constraint from $R_2$, we repeated the process described in Sec.~\ref{sec: R2 calculations} at regular intervals between $m_{\mathrm{DP}}=1$~keV, where the bound is out-competed by that of $R$ and $10$~keV, as dark photons more massive than this are produced too late in the E-AGB to significantly reduce its duration, and are already constrained by the RGB-tip luminosity. The resulting limit is shown in purple in Fig.~\ref{fig: R2 bound}.

This limit has two clear features. The first is a stalactite-like protrusion between $2$ and $3$~keV, where the bound is strongest. The second is another, broader stalactite to its right.

The latter coincides with dark photons which are not produced at all during the HB, and consequently is a direct limit on novel energy-loss during the E-AGB phase. This also explains its similarity to the RGB constraint of Fig.~\ref{fig: RGB-bound}, as such stars have similar plasma frequency profiles. In contrast with $R_2$, $L_{\mathrm{RGBT}}$ remains sensitive to dark photons which are only resonantly produced very late in that evolutionary phase, and hence it can be used to constrain more massive dark photons than the former.

The left stalactite-like feature, however, is precisely the region where dark photons resonantly produced in or near the HB core can seed dark photon core breathing pulses. As mentioned in the discussion of the middle panel of Fig.~\ref{fig: R2 distributions}, the dual elongation and shortening of the HB and AGB due to long, repeated CBPs, compounded by strong novel energy-loss during the E-AGB itself, makes $R_2$ particularly sensitive in this region, leading to a strong bound.

\subsection{Systematic uncertainties}
\label{sec: R/R2 systematics}

In App.~\ref{app: systematics}, we investigate the impact on $R$ and $R_2$ of two important sources of systematic uncertainty, namely deviation from our adopted averaged-Gaunt factors (Eq.~\ref{eq: Gbar wo screening}) and the treatment of mixing across the horizontal branch convective core boundary.

Regarding the former, we find that our limits are insensitive to large variations in $\Bar{g}_{ff}$, except when dark photon core breathing pulses are responsible for the constraint. In this scenario, corrections to $\Bar{g}_{ff}$ from special relativity, partial degeneracy and screening largely cancel one another, meaning deviations from Eq.~\ref{eq: Gbar wo screening} are small and our result is accurate.

It is widely acknowledged in the astrophysical literature that the treatment of mixing across the convective core boundary constitutes one of the most important sources of systematic uncertainty on theoretical calculations of $R$ and $R_2$ \cite{Lattanzio2, Lattanzio3}. If the method for modelling this mixing is altered, it typically systematically increases one of $R$ or $R_2$, while decreasing the other. Consequently, either limits derived from $R$ strengthen and those from $R_2$ weaken, or vice versa.

In our previous work \cite{Dolan:2022kul}, we were able to use this fact to mitigate this uncertainty by identifying the scheme which led to the most conservative overall constraint on the axion-photon coupling. For dark photons, as the constraints from $R$ and $R_2$ have different regions of parameter-space in which they are dominant, this approach is no longer possible. Instead, we investigated the variation of the limits presented in this section for a number of benchmark dark photon masses, when the scheme for modelling convective mixing is changed.

The results of our analysis are in agreement with our previous results, i.e.\ that changing mixing scheme either strengthens the limit from $R_2$ and weakens that from $R$, or vice versa. There is one exception to this, for a mixing scheme which limits the rate of helium influx into the horizontal branch core. This scheme reduces the size and duration of core breathing pulses, thereby reducing the strength of our limit from $R_2$ in the vicinity of $m_{\rm{DP}}\approx2$~keV by a factor of 1.23. Despite this, dark photon core breathing pulses persist in all investigated schemes, including those which remove their astrophysical counterparts. 

\section{Conclusion}
\label{sec: conclusion}

We have undertaken dynamic, self-consistent stellar evolution simulations to study the impact of dark photons on globular cluster populations. 
As observables we have used the red-giant branch (RGB) tip luminosity, and the parameters $R$ and $R_2$, which are the ratios of horizontal branch (HB) to RGB ($R$), and asymptotic giant branch (AGB) to HB ($R_2$) stars in globular clusters. This allows us to provide improved bounds on the dark photon parameter space.

For the RGB-tip luminosity we found that static stellar models miss the resonant production of dark photons with masses less than the assumed core temperature. In this region we obtained stronger bounds on the mixing parameter $\chi$ by a factor of approximately 5. On the other hand, static models overstate the strength of bounds for masses higher than the assumed core temperature, since the production of dark photons occurs so late in the RGB evolution that dark photon energy-loss does not have time to impact the RGB tip luminosity in an observable way. 

Our limit from $R$ closely matches that of Ref.~\cite{2016JCAP...05..057G}, except for $m_{\mathrm{DP}}\lesssim 1$~keV. In this case, resonant dark photon production during the RGB decreases the duration of this evolutionary phase causing our constraint to relax. In most circumstances, this relaxation occurs for values of $\chi$ already excluded by the RGB-tip luminosity constraint, leaving our overall bound unaffected. The notable exception to this is for $m_{\mathrm{DP}}=1$~keV.

When computing $R$ and $R_2$, we found that resonant dark photon production near the boundary of the convective core during the late stages of horizontal branch evolution seeded large convective episodes, which we term \textit{dark photon core breathing pulses}, which delayed the exhaustion of central helium. The $R_2$-parameter was particularly effective at constraining dark photons in the region of parameter space where these are relevant ($m_{\mathrm{DP}}\approx2.51$~eV), leading to an improvement on the bounds on $\chi$ by half an order of magnitude.

While dark photons with masses greater than about $3.98$~keV were too massive to be resonantly produced in HB stars, their impact on the AGB evolutionary phase nevertheless enabled them to be constrained by $R_2$. We found that this limit had comparable constraining power to that from the RGB-tip luminosity for masses up to $10$~keV.

There are also direct detection constraints from various experiments in this region of parameter space. However, they assume that the dark photons constitute the entirety of the cosmological relic abundance. Our constraints have the advantage of being entirely independent of the relic density of dark photons, since they are produced inside the stars we consider in a calculable way and our observables rely on stellar energy-loss.

Our work underscores the importance of dedicated stellar simulations to provide constraints on Beyond the Standard Model physics. In light of this, it would be interesting to revisit stellar limits on other light vector bosons, such as $U(1)_{B-L}$ and $U(1)_{L_i-L_j}$~\cite{Foot:1990mn,He:1990pn,He:1991qd}.


\acknowledgments
This work was supported in part by the Australian Research Council through the ARC Centre of Excellence for Dark Matter Particle Physics, CE200100008 and the Australian Government Research Training Program Scholarship initiative.

\appendix

\section{Including dark photon energy-loss in \texttt{MESA}}
\label{app: DP energy-loss}
Novel sources of energy-loss can be included in \texttt{MESA} by adding terms to the \texttt{neu} module, which calculates the energy-loss to neutrino production in each cell of the model \cite{Friedland, Dolan:2022kul, Dolan2021}. 

Energy-loss via the longitudinal channel has previously been included in stellar evolution simulations \cite{Vinyoles2015:10015V, Ayala:2014pea}, including in \texttt{MESA} itself. These typically make use of the resonant production approximation (Eq.~\ref{eq: eps L}) and set energy-loss to zero in stellar regions with plasma frequencies smaller than $m_{\mathrm{DP}}$. This approximation is valid so long as one is considering a region of parameter space in which off-resonant longitudinal dark photon production is negligible. As we are primarily concerned with the effects of transverse dark photon production, we also take this approximation.

No such simple approximation exists for the transverse production channel, and therefore we must include energy-loss of the form given by Eqs.~\ref{eq: energy-loss}, \ref{eq: T production} and \ref{eq: GammaT} in our stellar models. We can express $\epsilon_{\rm{T}}$ as
\begin{equation}
    \label{eq: epsilon interp form}
    \epsilon_{\rm{T}}=\frac{\chi^2m_{\mathrm{DP}}^4}{\pi^2\rho}G(m_{\mathrm{DP}}, T, \omega_{\mathrm{pl}},  \Gamma_1) \,,
\end{equation}
where we define the function $G$ to be
\begin{equation}
\begin{split}
\centering
    G(m_{\mathrm{DP}}, T, \omega_{\mathrm{pl}} &, \Gamma_1) = \int_{m_{\mathrm{DP}}}^{\infty} \frac{d\omega \omega^2 \sqrt{\omega^2-m_{\mathrm{DP}}^2}}{e^{\omega/T}-1}\\ & \times \frac{\Gamma_{\rm{T}}}{(m_{\mathrm{DP}}^2-\omega_{\mathrm{pl}}^2)+(\omega\Gamma_{\rm{T}})^2}.
\end{split}
\end{equation}
Once again, $m_{\mathrm{DP}}$, $T$ and $\omega_{\mathrm{pl}}$ are the dark photon mass, local stellar temperature and plasma frequency respectively. The parameter $\Gamma_1$ arises from our parametrisation of the plasmon damping rate (Eq.~\ref{eq: GammaT})
\begin{equation}
    \label{eq: parametrised plasmon decay rate}
    \Gamma_{\rm{T}}(T, \omega_{\mathrm{pl}}, \Gamma_1) = \Gamma_1\frac{(1-e^{\omega/T})\Bar{g}_{ff}(\omega/T)}{\omega^3}+\frac{2\alpha}{3m_e}\omega_{\mathrm{pl}}^2 \,.
\end{equation}
The issue of computing $\epsilon_{\rm{T}}$ within \texttt{MESA} simulations therefore reduces to the ability to swiftly and accurately evaluate the function $G$.

\tocless\subsection{Methods}
In our previous works novel energy-loss (e.g.\ to ALPs) was included in \texttt{MESA} by interpolating over a pre-processed 2D grid \cite{Dolan2021, Dolan:2022kul}. However, as our function $G$ depends on four variables, a more considered approach must be taken.

The data used in this process are structured in the following manner.
\begin{itemize}
    \item At the highest level, data are organised into mass directories, each of which assumes a different value of $\log\, m_{\mathrm{DP}}$.
    \item The next level below contains 61 temperature directories, which span values of $\log\, T$ between 6.00 and 10.00 in intervals of 0.05.
    \item Within each temperature directory is a grid of 2D tables, each of which corresponds to different ranges of $\log\, \omega_{\mathrm{pl}}$ and $\log\, \Gamma_1$ values. The latter are spaced linearly, while the former are densely clumped near the point where $\omega_{\mathrm{pl}}=m_{\mathrm{DP}}$ to enable better resolution of the resonant peak.
\end{itemize}

As the value of $m_{\mathrm{DP}}$ does not change throughout a stellar lifetime, it can be fixed at the start of each simulation by directing the code to only consider data within the appropriate mass directory. The process for evaluating $G$ in each cell is then:
\begin{enumerate}
    \item Given the cell temperature $T$, locate the two closest temperature directories $T_1$, $T_2$.
    \item Determine cell values of $\omega_{\mathrm{pl}}$ and $\Gamma_1$. Interpolate over the corresponding table in each temperature directory to find $G_1=G(m_{\mathrm{DP}}, T_1, \omega_{\mathrm{pl}}, \Gamma_1)$ and $G_2=G(m_{\mathrm{DP}}, T_2, \omega_{\mathrm{pl}}, \Gamma_1)$.
    \item Interpolate between $G_1$ and $G_2$ to find an estimate for $G=G(m_{\mathrm{DP}}, T, \omega_{\mathrm{pl}}, \Gamma_1)$, which is then fed into Eq.~\ref{eq: epsilon interp form} to evaluate $\epsilon_{\rm{T}}$.
\end{enumerate}

\tocless\subsection{Other considerations}

One key step in accurately implementing energy-loss to transverse dark photons in \texttt{MESA} was to ensure our simulations had sufficient spatial resolution to resolve the resonant production region where $\epsilon_{\rm{T}}$ peaks. To accomplish this, we utilised the \texttt{run\_star\_extras} routine \texttt{other\_mesh\_delta\_coeff\_factor}, which allows one to set a custom resolution in stellar regions obeying user-specified properties. Our custom \texttt{mesh\_delta\_coeff} value was set to 0.01 in regions where $\omega_{\mathrm{pl}}$ was within 1\% of $m_{\mathrm{DP}}$, which proved sufficient.

\vspace{1cm}
\section{Systematic uncertainties}
\label{app: systematics}
In this section we describe the impact on our constraints of two sources of systematic uncertainty: (i) the treatment of mixing across convective boundaries and (ii) inaccuracies in calculating $\Gamma_{\rm{T}}$ in Eq.~\ref{eq: GammaT}.

\tocless\subsection{Mixing across convective core boundaries}
\label{sec: CC boundaries}

\begin{table*}
\centering
\begin{tabular}{ccccccc}
\hline\hline
\multirow{2}{*}{\textbf{Scheme}} & \multirow{2}{*}{\textbf{Parameters}}                                                                                       & \multicolumn{2}{c}{$R$}                         & \multicolumn{3}{c}{$R_2$}                                                 \\
                        &                                                                                                                   & $m_{\mathrm{DP}}=0.79$~keV & $m_{\mathrm{DP}}=2.51$~keV & $m_{\mathrm{DP}}=1.00$~keV & $m_{\mathrm{DP}}=2.51$~keV & $m_{\mathrm{DP}}=3.98$~keV  \\
\hline
SO                      & $f_{\mathrm{ov}}=0.01$                                                                                            & 1.10                   & 1.10                   & 0.66                   & <0.72                   & 0.66                    \\
SC                      & $f_{\mathrm{ov}}=0.001$~and~$\alpha_{\mathrm{SC}}=0.1$~                                                           & 1.12                   & 0.89                   & 0.81                   & <0.72                   & 0.79                    \\
SC                      & $f_{\mathrm{ov}}=0.001$ and $\alpha_{\mathrm{SC}}=0.01$ & 1.10                   & 0.93                   & 0.95                   & <0.72                   & 0.79                    \\
PM                      & $\alpha_i=0.42$                                                                                                   & 1.00                   & 0.74                   & 0.89                   & 1.02                   & 0.51                    \\
PM                      & $\alpha_i=0.21$                                                                                                   & 1.02                   & 0.71                   & 1.00                   & 1.23                   & <0.5                     \\
PM                      & $\alpha_i=0.042$                                                                                                  & 1.00                   & 0.62                   & 1.23                   & 0.83                   & <0.5                     \\
\hline\hline
\end{tabular}
\caption{Revised $R$ and $R_2$ constraints for benchmark dark photon masses when other implementations of convective boundary schemes are implemented. Each entry in the table corresponds to the limiting value of $\chi$ given the implementation specified in the first two columns, normalised to the equivalent limit from Sec.~\ref{sec: star cluster counts}. Entries $<1$ indicate the new limit is stronger than the old.}
\label{tab: CM}
\end{table*}


The treatment of mixing across the horizontal branch convective core boundary is one of the major sources of systematic and stochastic uncertainty on theoretical predictions for $R$ and $R_2$. The mechanisms behind this mixing and its impact on axion bounds derived from these two globular cluster parameters were discussed comprehensively in our previous paper \cite{Dolan:2022kul} (specifically in Sec. 2 and App. C). Therefore, we shall repeat only the salient points here before detailing the effect of this source of uncertainty on the limits from Sec.~\ref{sec: RGB tip luminosity} and ~\ref{sec: star cluster counts}.

A key concept for the forthcoming discussion is the phenomenon of \textit{convective overshoot}, i.e.\ the inertial tendency for convective elements to push into the stable region beyond the convective boundary, thereby eliciting some form of chemical mixing across the boundary. As was discussed in Sec.~\ref{sec: The HB}, when the products of helium-burning are progressively transferred across the convective boundary, episodes of core growth and splitting result. The inner structure of HB stars can therefore be described in terms of an inner core which is always convective, an outer helium-rich zone which is never touched by the convective core and an intermediate region between the two which is partially mixed by these episodes. 

Schemes for modelling mixing across convective boundaries in stellar evolution codes can be distinguished based on how they model mixing: (i) at the convective boundary and (ii) in the aforementioned intermediate region. \texttt{MESA} is furnished with a number of different convective boundary schemes, which we have previously categorised based on their different approaches to the first point~\cite{Dolan:2022kul}. 

%
\tocless\subsubsection{Exponential overshoot approaches}
Exponential overshoot approaches model mixing across the convective boundary as a diffusive process. Specifically, the diffusion coefficients in the model exponentially decrease the further one looks from the convective boundary. A free parameter $f_{\mathrm{ov}}$ is introduced, which sets the distance scale for exponential decrease. Larger values of $f_{\rm{ov}}$ correspond to stronger overshoot.

We consider two overshoot based approaches in this section. These are:
\begin{enumerate}
    \item \textit{Standard overshoot} (SO), which uses exponential overshoot with no additional scheme for modelling mixing in the aforementioned intermediate region.
    \item \textit{Semiconvection} (SC), which also uses exponential overshoot and a \textit{semiconvective} mixing scheme in the intermediate region. The rate of mixing of the latter is set by another free parameter $\alpha_{\mathrm{SC}}$, larger values of which correspond to more efficient mixing.
\end{enumerate}

\tocless\subsubsection{Instantaneous mixing}
An alternative to exponential overshoot is \texttt{MESA}'s instantaneous mixing approach. This operates by picking a candidate cell for the convective core boundary and checking whether the cell immediately next to this one becomes convective if its contents were mixed with the rest of the core. If so, this cell is considered as the new boundary candidate. This process repeats until a candidate is found whose neighbour remains stable after this mixing, at which point the candidate is designated as the convective core boundary.

Of the two instantaneous mixing schemes available in \texttt{MESA}, we consider \textit{predictive mixing} (PM) in this section, which is partially based on the maximal overshoot scheme of Ref.~\cite{Lattanzio1}. In addition to this, predictive mixing contains the option to reduce the frequency and impact of core breathing pulses by limiting the rate of helium ingestion into the core, so-called \textit{Spruit overshoot} \cite{Lattanzio3, Spruit2015}. The rate of ingestion is controlled by free parameter $\alpha_i$, where low values correspond to more restricted ingestion.

\vspace{1cm}
\tocless\subsubsection{Impact on \texorpdfstring{$R$}{R} and \texorpdfstring{$R_2$}{R2} limits}

In our previous work, we found that varying the adopted mixing scheme and free parameters had the opposite effect on our limits from $R$ and $R_2$, i.e.\ they would strengthen our limit from $R$ and weaken that from $R_2$ or vice versa \cite{Dolan:2022kul}. Given that only the axion-photon coupling was varied, it was therefore possible to marginalise across many different implementations of the available schemes to derive a robust limit.

In this work, as energy-loss to dark photons depends on both their mass and kinetic mixing, the $R$ and $R_2$ limits from Sec.~\ref{sec: star cluster counts} are dominant in different regions of parameter space, meaning that this approach is no longer possible. Therefore, instead of computing a single combined limit which we know to be the most robust, we shall simply characterise the degree to which these bounds vary when different implementations of the available mixing schemes are used. As a full parameter-scan is computationally intense, we conduct this for specific benchmark values of $m_{\mathrm{DP}}$.

The limits presented in Sec.~\ref{sec: star cluster counts} were computed using the standard overshoot scheme with $f_{\mathrm{ov}}=0.001$. Here we contrast these results with those obtained given:
\begin{itemize}
    \item Standard overshoot (SO) with $f_{\mathrm{ov}}=0.01$,
    \item Semiconvection (SC) with $f_{\mathrm{ov}}=0.001$ and $\alpha_{\mathrm{SC}}=0.1$ and $0.01$,
    \item Predictive mixing (PM) with $\alpha_i=0.42$, $\alpha_i=0.21$ and $\alpha_i=0.042$,
\end{itemize}
which cover a wide range of the values found in the literature \cite{Lattanzio2, Lattanzio3, MIST1}.

\begin{table}
\centering
\begin{tabular}{c|cccc} 
\hline\hline
\multirow{2}{*}{$f_g$} & \multicolumn{4}{c}{$m_{\rm{DP}}$}                                                                       \\
                       & $1$~keV & $6.31$~keV & $15.8$~keV & $20.0$~keV  \\ 
\hline
10                     & 0.987                 & 1.006                    & 0.333                    & 0.329                     \\
0.1                    & 0.982                 & 1.006                    & 2.760                    & 2.256                     \\
0.01                   & 0.984                 & 1.004                    & 4.259                    & 2.912                     \\
0.003                  & -                     & -                        & 4.476                    & 3.027                     \\
\hline\hline
\end{tabular}

\caption{Revised RGB-tip constraints for benchmark dark photon masses when the averaged-Gaunt factors are multiplied through by a constant factor $f_g$. Each entry in the main part of the table corresponds to the new limiting value of $\chi$ indicated in the first column, normalised to the equivalent limit from Sec.~\ref{sec: RGB tip luminosity}. Entries $<1$ indicate the new limit is stronger than the old.}
\label{tab: Gaunt RGBT results}
\end{table}

Using these schemes we calculate revised limits for $R$ at $m_{\mathrm{DP}}=0.79$~keV, where $R$ is the leading constraint, and $m_{\mathrm{DP}}=2.51$~keV, which coincides with a resonant production region in the HB core. For $R_2$, our benchmark dark photon masses are $m_{\mathrm{DP}}=1$~keV, which is the lowest mass for which we have a constraint, $m_{\mathrm{DP}}=2.51$~keV for the same reasons as $R$ and $m_{\mathrm{DP}}=3.98$~keV, where resonant production only occurs in AGB stars.

The exclusion limits given these new schemes are then normalised to the equivalent values from Sec.~\ref{sec: star cluster counts}. These are listed in Tab.~\ref{tab: CM}. Values less than one indicate the new limit is stronger than that given standard overshoot with $f_{\mathrm{ov}}=0.001$.

The simplest case we consider is when $f_{\mathrm{ov}}$ is increased to 0.01. In this scenario, increasing the distance scale over which overshoot occurs results in systematically larger HB durations. This leads to weaker bounds from $R$ and stronger $R_2$ limits, as identified in Ref.~\cite{Dolan:2022kul}.

When a semiconvective scheme is used, the same general pattern is observed, i.e.\ that our limit weakens for $m_{\mathrm{DP}}\lesssim1$~keV when $R$ is dominant over $R_2$, but strengthens in both the regions where DP-CBPs occur ($m_{\mathrm{DP}}=2.51$~keV) as well as when transverse dark photons are only produced during the AGB ($m_{\mathrm{DP}}\gtrsim3.98$~keV). We note that this is not the same pattern that we observed in our previous paper, where a semiconvective scheme led to the most conservative overall limit on $R_2$.

Finally, if the predictive mixing scheme is adopted, we can limit the influx of helium into the convective core during the HB phase, thereby mitigating the impact of dark photon core breathing pulses. For the predictive mixing parameter $\alpha_i=0.42$, the limit is approximately consistent with that from Sec.~\ref{sec: star cluster counts} in the region where DP-CBPs occur. If this is decreased to $\alpha_i=0.21$, i.e.\ the rate of helium ingestion is slowed, the limit weakens. We note, however, that dark photon core breathing pulses still occur in these simulations, albeit with reduced frequency and magnitude. Finally, when $\alpha_i=0.042$, multiple very small core breathing pulses occur which result in a limit which is ultimately stronger than our original one.

On the other hand, our limit from $R_2$ can increase by a factor of two or more for $m_{\mathrm{DP}}\gtrsim3.98$~keV when predictive mixing is used, and the bound from $R$ for $m_{\mathrm{DP}}\lesssim1$~keV is consistent with that from Sec.~\ref{sec: star cluster counts}.

In summary, varying the adopted scheme and free parameters for modelling mixing across convective boundaries can cause our limit to pivot, strengthening the constraint for certain dark photon masses, while weakening it for others. We note that the results in Sec.~\ref{sec: star cluster counts}, generated using standard overshoot with $f_{\rm{ov}}=0.001$, stand to improve in the mass range $m_{\rm{DP}}\gtrsim3.98$~keV in all of the scenarios examined. Furthermore, dark photon core breathing pluses persist in some form throughout all convective boundary schemes we consider, even those which limit the rate of helium ingestion into the core.

\tocless\subsection{Uncertainty in $\epsilon_{\rm{T}}$}
\label{sec: Uncertainty in EpsT}

Throughout this work we have assumed that, because our new limits are driven by energy-loss to resonantly produced transverse dark photons, they do not depend sensitively on the details of $\Gamma_{\rm{T}}$. In this section we test the validity of this assumption.

We approach this in the simplest possible way: by calculating how our limits from $L_{\text{RGBT}}$, $R$ and $R_2$ differ at benchmark values of $m_{\mathrm{DP}}$ when $\Bar{g}_{ff}$ is multiplied by a constant factor $f_g$.

We can crudely estimate the effect of neglecting contributions to $\Bar{g}_{ff}$ in $\Gamma_{\rm{T}}$ from electron degeneracy, screening and relativistic corrections by computing the \textit{total} Gaunt factor
\begin{equation}
    \langle \Bar{g}_{ff}^{\text{FD}}\rangle(\eta,\, \gamma^2)=\int_{m_{\mathrm{DP}}/T}^{\infty}e^{-u}\Bar{g}^{\text{FD}}_{ff}(u,\, \eta,\, \gamma^2)du,
\end{equation}
where $\Bar{g}^{\text{FD}}_{ff}$ are the relativistic Fermi-Dirac-averaged Gaunt factors of~\cite{1987ApJS...63..661N} corrected for screening using the Green approximation \cite{RM-2580-AEC} (see Ref.~\cite{ARMSTRONG201461} for further details). The factors $u$ and $\eta$ were defined in Sec.~\ref{sec: off-resonant production} and the dimensionless parameter $\gamma^2$ is given by $\gamma^2=\frac{Z^2\text{Ry}}{T}$, where $\text{Ry}$ is the Rydberg energy. We note that, ordinarily the total Gaunt factor is obtained by integrating over all photon energies, however, as Eq.~\ref{eq: energy-loss} has $m_{\mathrm{DP}}$ as its lower integration terminal, we adopt this here too.

An approximate value for $f_g$ can then be obtained by evaluating the ratio $\langle \Bar{g}_{ff}^{\mathrm{FD}}\rangle/\langle \Bar{g}_{ff} \rangle$, where $\langle \Bar{g}_{ff}\rangle$ is the equivalent total Gaunt factor when Eq.~\ref{eq: Gbar wo screening} is used.
\vspace{1cm}

\tocless\subsubsection{\texorpdfstring{$L_{\rm{RGBT}}$}{LRGBT} limit}

\begin{table}
\centering
\begin{tabular}{c|cc} 
\hline\hline
\multirow{2}{*}{$f_g$} & \multicolumn{2}{c}{$m_{\rm{DP}}$}  \\
                       & $0.79$~keV & $2.51$~keV            \\ 
\hline
10                     & 1.001    & 1.012                       \\
0.1                    & 1.000    & 1.012                           \\
\hline\hline
\end{tabular}
\caption{Revised $R$-parameter constraints for benchmark dark photon masses when the averaged-Gaunt factors are multiplied through by a constant factor $f_g$. Each entry in the table corresponds to the new limiting value of $\chi$ indicated in the first column, normalised to the equivalent limit from Sec.~\ref{sec: RGB tip luminosity}. Entries $<1$ indicate the new limit is stronger than the old.}
\label{tab: R Gaunt}
\end{table}

For our limit from $L_{\mathrm{RGBT}}$, we pick benchmark values of $m_{\mathrm{DP}}=1$~keV, $6.31$~keV, $15.8$~keV and $20.0$~keV to test the validity of the assumption that transverse dark photon production does not sensitively depend on $\Gamma_{\rm{T}}$. The first value corresponds to the lower end of dark photon masses considered, the second to the region where the RGB-tip bound is dominant and latter two to the parameter-space in which off-resonant production is responsible for the limit. In Tab.~\ref{tab: Gaunt RGBT results} we show the results of this process for $f_g=10$, $0.1$, $0.01$ and $0.003$. The entries in this table correspond to the new limiting values of the mixing parameter $\chi_{\mathrm{new}}(f_g)$ normalised to the corresponding value from Sec.~\ref{sec: RGB tip luminosity}. Values less than 1 indicate the new limit is stronger than the original.

For 1~keV and 6.31~keV dark photons, for which resonant transverse production is responsible for our limit, our results do not vary significantly over three orders of magnitude in $f_g$, which confirms the insensitivity of our limits to changes in $\Gamma_{\rm{T}}$ of this scale.

When $m_{\rm{DP}}=15.8$~keV and $20$~keV, RGB stars no longer obtain resonant production regions during their evolution meaning these limits arise from off-resonant production. Consequently, the limiting values of $\chi$ in this scenario do not enjoy the same protection against variation in $\Gamma_{\rm{T}}$ and hence show strong sensitivity to the adopted value of $f_g$.

Far from resonance, Eq.~\ref{eq: GammaT} (and hence $\epsilon_{\rm{T}}$) scales linearly with $\Gamma_{\rm{T}}$, meaning our new limits should scale as $\chi_{\rm{new}}\propto1/\sqrt{f_g}$. This is exactly what we see for the $f_g=10$ case. The $f_g=0.1$, $0.01$ and $0.003$ cases, however, deviate from this expectation. This occurs as diluting the strength of the free-free absorption term in $\Gamma_{\rm{T}}$ increases the comparative importance of Thomson scattering, particularly for large dark photon masses. If similar multiplicative factors are introduced for the Thomson scattering term, these entries in Tab.~\ref{tab: Gaunt RGBT results} increase to 3.11, 10.4 and 20.1 for $m_{\rm{DP}}=15.8$~keV and 3.04, 9.65 and 16.9 for $m_{\rm{DP}}=20.0$~keV respectively.

If we assume that the majority of energy-loss responsible for these off-resonant bounds occurs near the end of the RGB, when the central plasma frequency is largest, we can estimate an appropriate value of $f_g$ using standard conditions of an RGB-tip star (i.e. $T=10^8$~K, $\rho=2\times10^5\ \rm{g}~cm^{-3}$ and $\eta\sim20$). The corresponding value of $f_g$ are $0.005$ and $0.006$ for $m_{\rm{DP}}=15.8$~keV and $20$~keV respectively. We therefore conservatively set adjusted limits at $\chi_{\rm{new}}=14.1\chi_{\rm{old}}$ and $12.9\chi_{\rm{old}}$ for these dark photon masses, which feature in Fig.~\ref{fig: DP parameter space}.

\vspace{1cm}
\tocless\subsubsection{\texorpdfstring{$R$}{R}- and \texorpdfstring{$R_2$}{R2}-parameter limits}

\begin{table}
\centering
\begin{tabular}{c|ccc} 
\hline\hline
\multirow{2}{*}{$f_g$} & \multicolumn{3}{c}{$m_{\rm{DP}}$}  \\
                       & $1.00$~keV & $2.51$~keV & $3.98$~keV                 \\ 
\hline
10                     & 1.007    & 1.679    & 1.005                    \\
0.1                    & 0.991    & 0.914    & 1.002                    \\
\hline\hline
\end{tabular}
\caption{Revised $R_2$ constraints for benchmark dark photon masses when the averaged-Gaunt factors are multiplied through by a constant factor $f_g$. Each entry in the table corresponds to the new limiting value of $\chi$ indicated in the first column, normalised to the equivalent limit from Sec.~\ref{sec: RGB tip luminosity}. Entries $<1$ indicate the new limit is stronger than the old.}
\label{tab: R2 Gaunt}
\end{table}

To gauge the impact of varying $f_g$ on $R$ and $R_2$, we use the same benchmark masses which were examined in Sec.~\ref{sec: CC boundaries}, i.e.\ $m_{\mathrm{DP}}=0.79$~keV and $2.51$~keV for $R$ and $1$~keV, $2.51$~keV and $3.81$~keV for $R_2$. These were tested for $f_g=0.1$ and $10$, respectively, with the ratios of new to old constraints displayed in Tab.~\ref{tab: R Gaunt} and \ref{tab: R2 Gaunt} respectively.

Assuming standard HB core conditions ($T=10^8$~K, $\rho=10^4\ \rm{g}~cm^{-3}$ and $\eta\sim0.5$), we estimate $f_g$ to be $0.78$ and $0.84$ for the $m_{\rm{DP}}=0.79$~keV and $2.51$~keV cases respectively. Our limit from $R$ is clearly robust against changes in $\Gamma_{\rm{T}}$ of this scale.

The results for $R_2$ are more interesting. Clearly our limit for $m_{\rm{DP}}=2.51$~keV, where dark photon core breathing pulses are responsible for the bound, shows some sensitivity to the adopted values of $\Bar{g}_{ff}$. Specifically, for $f_g=10$, resonant production is less intense but spread over a wider region, which decreases the occurrence of DP-CBPs (see App.~\ref{app: HB CBPs}) weakening our limit. On the other hand, values of $f_g<1$ sharpen the resonant peak and increase both the rate of these core breathing pulses and the strength of our constraint. Taking values for the local temperature, density and degeneracy at the RPR in the late stages of HB evolution, we estimate $f_g$ to be 1.32 and 1.38 for carbon and oxygen respectively. This corresponds to a reduction in the strength of our constraint by a factor of 1.07.

Our other test-cases for the $R_2$ limit do not appear to be sensitive to the adopted value of $f_g$ and should therefore be considered robust.

In summary, either Eq.~\ref{eq: Gbar wo screening} is suitably accurate in scenarios where a large deviation from it could significantly affect our results, or our limits are insensitive to variation in $\Bar{g}_{ff}$ when it departs strongly from Eq.~\ref{eq: Gbar wo screening}. The sole exception to this occurs for the RGB-tip constraint for $m_{\rm{DP}}\gtrsim15.8$~keV, where only off-resonant dark photon production can give rise to a constraint. We have adjusted our limit in this region of parameter space in accordance with this.

\vspace{1cm}
\section{Core breathing pulses}
\label{app: HB CBPs}

In this appendix we describe in more detail the differences between standard astrophysical core breathing pulses and those which occur when resonant dark photon energy-loss occurs in the vicinity of the horizontal branch core.

\tocless\subsection{Convective instability in stars}

Understanding the distinction between astrophysical- and dark photon-core breathing pulses necessitates a discussion of the definition of the convective regions and boundaries in stellar models.

Formally, a stellar region will be stable against convection if it locally obeys the \textit{Schwarzschild criterion}:
\begin{equation}
    \nabla_{\rm{rad}}/\nabla_{\rm{ad}}<1,
\end{equation}
where $\nabla_{\rm{rad}}$ and $\nabla_{\rm{ad}}$ are the radiation and adiabatic temperature gradients respectively. These describe the temperature gradients which would be observed in stars if only the bulk movement of radiation or the adiabatic motion of convective elements were responsible for energy transport within them. They are respectively defined by:
\begin{equation}
    \label{eq: radiative temperature gradient}
    \nabla_{\mathrm{rad}}\equiv\bigg(\frac{\partial \ln T}{\partial \ln P}\bigg)_{\mathrm{rad}}=\frac{3}{16\pi a_R G}\frac{\kappa l P}{mT^4},
\end{equation}
and
\begin{equation}
    \label{eq: adiabatic temperature gradient}
    \nabla_{\mathrm{ad}}\equiv\bigg(\frac{\partial \ln T}{\partial \ln P}\bigg)_{\mathrm{ad}},
\end{equation}
where $a_R$ and G are the radiation density and gravitational constants, and $\kappa$, $l$, $P$, $m$ and $T$ are respectively the opacity, local luminosity, pressure, enclosed mass and temperature at some radial location. Typically Eq.~\ref{eq: adiabatic temperature gradient} is evaluated within the framework of mixing length theory in stellar models \cite{KippenhahnWeigert}. The convective core boundary is defined as the stellar region where the Schwarzschild criterion is broken, i.e. where $\nabla_{\rm{rad}}=\nabla_{\mathrm{ad}}$.

\tocless\subsection{The HeB convective core boundary}

As was briefly mentioned in Sec.~\ref{sec: The HB}, the boundary of the convective core is unstable during the HB. This instability arises due to the gradual production of carbon and oxygen in the core in conjunction with mixing across the convective core boundary.

The production of carbon and oxygen in the core drives up $\nabla_{\rm{rad}}$ via an increase in the local opacity. Convective overshoot then transports some of this more opaque mixture across the convective boundary, which increases the local value of $\nabla_{\mathrm{rad}}$ until the Schwarzschild criterion is broken and the region becomes unstable. This core growth, in turn, mixes the helium from the formerly stable region around the entire core, gradually shifting the $\nabla_{\mathrm{rad}}$ profile downwards again. 

In horizontal branch cores, the minima of $\nabla_{\rm{rad}}$ profiles can be at macroscopic distances from their convective boundaries. Consequently, if enough helium is introduced into the core, these locations will be the first to reach convective stability (i.e\ $\nabla_{\rm{rad}}/\nabla_{\rm{ad}}=1$) and the core splits into an inner region, which is still convective, and and outer region where the mixing is uncertain.

This process of core growth and splitting repeats many times during the HB on a timescale set by the temporal resolution of the model, and amounts to the instability of the core boundary. The key point to note is that changes in the core opacity profile ultimately drive this evolution, rather than any of the other parameters which appear in Eq.~\ref{eq: radiative temperature gradient}.\\

\tocless\subsection{Standard core breathing pulses}

\begin{figure}[t!]
    \centering
    \includegraphics[width=0.42\textwidth]{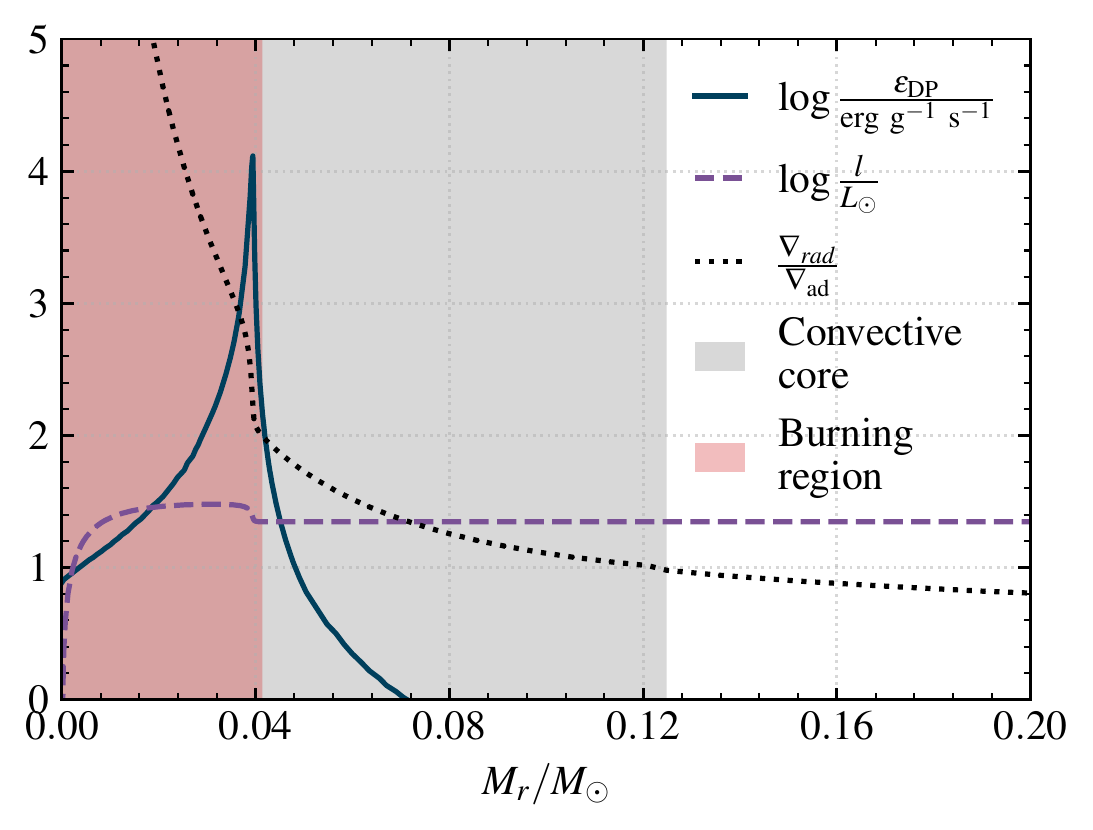}
    \caption{Stellar profiles showing the variation with radial mass coordinate of $\log\frac{\epsilon_{\rm{DP}}}{\rm{erg}~g^{-1}~s^{-1}}$ (solid, dark blue), $\log\frac{l}{L_{\odot}}$ (dashed, purple) and the ratio $\nabla_{\rm{rad}}/\nabla_{\rm{ad}}$ (dotted, black) over the innermost $0.2M_{\odot}$ of a horizontal branch star with initial mass $M_{\rm{init}}=0.82M_{\odot}$. The evolution of this star was simulated including energy-loss to $2.51$~keV dark photons. The red and grey regions indicate the radial extent of the burning and convective cores.}
    \label{fig: HB profile}
\end{figure}

The most dramatic result of the aforementioned convective boundary instability is astrophysical core breathing pulses. In the late stages of the horizontal branch phase, the major bottleneck for the triple-$\alpha$ and $^{12}C(\alpha,\gamma)^{16}O$ reaction rates is the supply of helium in the core. As the latter begins to dwindle, so do the reaction rates, elevating the core temperature and density profiles.

In these conditions, any influx of helium as a result of core growth is met with a rapid rise in nuclear-burning activity. This, in turn, encourages further convective excursions into the helium-rich region, until a suitably large amount of helium has been deposited into the core, which once again settles into quiescent helium-burning. Such events are typically more extreme the closer they occur to the TAHB, and extend HB duration by an amount that depends on the specific convective boundary scheme being implemented.

Crucially, the core growth which facilitates these core breathing pulses arises due to changes in the opacity profile from the production of helium-burning products.

\vspace{1cm}

\tocless\subsection{Dark photon core breathing pulses}

\begin{figure}[t]
    \centering
    \includegraphics[width=0.47\textwidth]{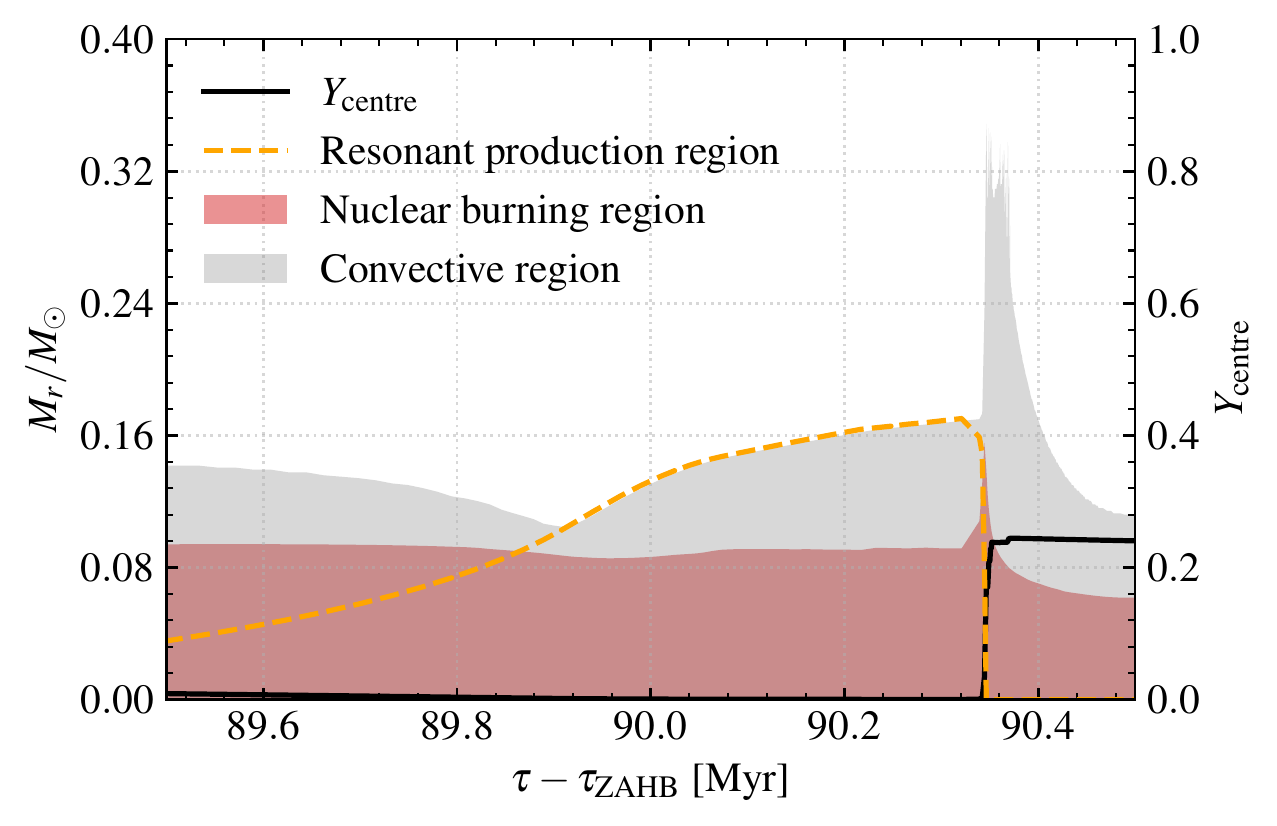}
    \caption{Zoomed version of the Kippenhahn diagram from the bottom right panel ($m_{\rm{DP}}=3.16$~keV and $\chi=10^{-15}$) of Fig.~\ref{fig: HB Kippenhahns}, emphasising the structural evolution in the build-up to the dark photon core breathing pulse at $\tau-\tau_{\rm{ZAHB}}\approx90.35$~Myr. The location of the dark photon resonant production region is indicated by the dashed orange line. The value of $Y_{\rm{centre}}$ is given by the dashed black line.}
    \label{fig: Zoomed Kippenhahn}
\end{figure}

By contrast, dark photon core breathing pulses occur due to changes in the local luminosity in the resonant production regions in conjunction with the outward movement of said region late in the horizontal branch phase.

To illustrate this, consider a $0.82M_{\odot}$ HB star which is experiencing energy-loss to dark photons with $m_{\mathrm{DP}}=2.5$~keV, which places the resonant production region within the convective core. The profile of the inner $0.2M_{\odot}$ of such a star is shown in Fig.~\ref{fig: HB profile}. The red region indicates values of $M_{r}$ which support helium-burning with $\epsilon_{\mathrm{He}}>50$~erg g$^{-1}$ s$^{-1}$, and the convective core is shown in light grey. We also show  the energy-loss rate per unit mass to dark photon production $\epsilon_{\mathrm{DP}}$ (dark blue solid), local luminosity (purple dashed) and the ratio $\nabla_{\mathrm{rad}}/\nabla_{\mathrm{ad}}$ (black dotted) as functions of the radial mass coordinate. The clear peak visible in $\epsilon_{\mathrm{DP}}$ corresponds to the resonant transverse production region. 

During the HB, nuclear burning peaks at the centre of the star producing a flux of photons which then slowly diffuse outward through the ionised and dense stellar interior, towards the exterior. On this journey, these photons must pass through a stellar region with plasma frequency equal to the hypothesised dark photon mass, at which point resonant photon to dark photon conversion can occur. This manifests in our models as a decrease in the local luminosity $l$, and hence $\nabla_{\mathrm{rad}}$ via Eq.~\ref{eq: radiative temperature gradient}, across the resonant production region.

In this scenario, the decrease in $\nabla_{\mathrm{rad}}$ is insufficient to take it below $\nabla_{\mathrm{ad}}$ and the region immediately above the resonant production region remains convective. In the late stages of HB evolution, however, as the RPR moves outward due to the increasing density profile of the star, the alternative scenario occurs in which the decrease in $\nabla_{\mathrm{rad}}$ across the RPR takes it below $\nabla_{\mathrm{ad}}$, and the Schwarzschild criterion is met. Hence, the location of the RPR now \textit{defines} the convective core boundary. The transition between these two regimes is shown in the Kippenhahn diagram in Fig.~\ref{fig: Zoomed Kippenhahn}, which is a zoomed version of case (iii) in Fig.~\ref{fig: HB Kippenhahns}.

As the RPR location continues to shift outward due to the contraction of the HB core, so does the convective core boundary. If this period of growth causes a stellar region with appreciable helium supply to become convective, its contents get mixed into a very hot, dense core which is starved of nuclear fuel. This sparks a short burst of intense nuclear burning, indicated by the red spike at $\tau-\tau_{\mathrm{ZAHB}}\approx90.35$~Myr, which precipitates a large convective excursion into the helium rich zone. As a result, a large amount of helium is deposited into the core ($Y_{\mathrm{centre}}\gtrsim0.2$ in this case), which settles back into quiescent helium-burning.

We note that, while the mechanism causing core growth is different in the standard and dark photon cases, the stellar response to the injection of helium into the core is qualitatively similar in both scenarios. For this reason we term the events described here \textit{dark photon core breathing pulses}.

Interestingly, as dark photon core breathing pulses can only occur in the very late stages of the HB phase when the stellar density profile increases substantially enough to facilitate the outward movement of the RPR, the magnitude of these events is always large. In contrast, standard core breathing pulses, which occur at the whim of opacity-induced core growth, may occur earlier in which case they can be small.

Finally, it was noted in Sec.~\ref{sec: Uncertainty in EpsT} that scaling the adopted values of the averaged-Gaunt factors $\Bar{g}_{ff}$ by a constant factor $f_g$ changed our calculated limit from $R_2$ for the $m_{\rm{DP}}=2.51$~keV case. This coincides with the region of parameter-space in which dark photon core breathing pulses are responsible for the bound. Specifically, it was found that, for $f_g=10$ our limit had to relax by a factor of 1.7, whereas if $f_g=0.1$, it improved by around 10\%.

This can be understood through consideration of Fig.~\ref{fig: HB profile}. When $f_g=10$, the profile of $\epsilon_{\rm{DP}}$ will be spread over a wider region, but peak less intensely, meaning the reduction in $\nabla_{\rm{rad}}/\nabla_{\rm{ad}}$ will occur more gradually. Consequently, in the final moments before a core breathing pulse is sparked, the convective core boundary will not be exactly at the RPR location, as in the above example, but will lag marginally behind it. It therefore has access to less helium than when $f_g=1$, making dark photon core breathing pulses less likely and causing the limit to weaken.

In the alternate scenario, where $f_g<1$, the decrease in $\nabla_{\mathrm{rad}}/\nabla_{\mathrm{ad}}$ becomes more step-like. The core boundary therefore exactly coincides with the location of the RPR, marginally improving our constraint.

\bibliographystyle{JHEP}
\bibliography{bibliography.bib}

\providecommand{\href}[2]{#2}\begingroup\raggedright\begin{thebibliography}{10}

\bibitem{Arias:2012az}
P.~Arias, D.~Cadamuro, M.~Goodsell, J.~Jaeckel, J.~Redondo and A.~Ringwald,
  \emph{{WISPy Cold Dark Matter}},
  \href{https://doi.org/10.1088/1475-7516/2012/06/013}{\emph{JCAP} {\bfseries
  06} (2012) 013} [\href{https://arxiv.org/abs/1201.5902}{{\ttfamily
  1201.5902}}].

\bibitem{An:2013yfc}
H.~An, M.~Pospelov and J.~Pradler, \emph{{New stellar constraints on dark
  photons}}, \href{https://doi.org/10.1016/j.physletb.2013.07.008}{\emph{Phys.
  Lett. B} {\bfseries 725} (2013) 190}
  [\href{https://arxiv.org/abs/1302.3884}{{\ttfamily 1302.3884}}].

\bibitem{2016JCAP...05..057G}
M.~{Giannotti}, I.~{Irastorza}, J.~{Redondo} and A.~{Ringwald}, \emph{{Cool
  WISPs for stellar cooling excesses}},
  \href{https://doi.org/10.1088/1475-7516/2016/05/057}{\emph{\jcap} {\bfseries
  2016} (2016) 057} [\href{https://arxiv.org/abs/1512.08108}{{\ttfamily
  1512.08108}}].

\bibitem{An:2020bxd}
H.~An, M.~Pospelov, J.~Pradler and A.~Ritz, \emph{{New limits on dark photons
  from solar emission and keV scale dark matter}},
  \href{https://doi.org/10.1103/PhysRevD.102.115022}{\emph{Phys. Rev. D}
  {\bfseries 102} (2020) 115022}
  [\href{https://arxiv.org/abs/2006.13929}{{\ttfamily 2006.13929}}].

\bibitem{Goodsell:2009xc}
M.~Goodsell, J.~Jaeckel, J.~Redondo and A.~Ringwald, \emph{{Naturally Light
  Hidden Photons in LARGE Volume String Compactifications}},
  \href{https://doi.org/10.1088/1126-6708/2009/11/027}{\emph{JHEP} {\bfseries
  11} (2009) 027} [\href{https://arxiv.org/abs/0909.0515}{{\ttfamily
  0909.0515}}].

\bibitem{Raffelt:1996wa}
G.G.~Raffelt, \emph{{Stars as laboratories for fundamental physics}: {The
  astrophysics of neutrinos, axions, and other weakly interacting particles}}
  (5, 1996).

\bibitem{Sieverding:2021jfa}
A.~Sieverding, E.~Rrapaj, G.~Guo and Y.Z.~Qian, \emph{{Impact of Dark Photon
  Emission on Massive Star Evolution and Pre-Supernova Neutrino Signal}},
  \href{https://doi.org/10.3847/1538-4357/abee84}{\emph{Astrophys. J.}
  {\bfseries 912} (2021) 13}
  [\href{https://arxiv.org/abs/2101.08672}{{\ttfamily 2101.08672}}].

\bibitem{Redondo:2013lna}
J.~Redondo and G.~Raffelt, \emph{{Solar constraints on hidden photons
  re-visited}},
  \href{https://doi.org/10.1088/1475-7516/2013/08/034}{\emph{JCAP} {\bfseries
  08} (2013) 034} [\href{https://arxiv.org/abs/1305.2920}{{\ttfamily
  1305.2920}}].

\bibitem{Vinyoles2015:10015V}
N.~{Vinyoles}, A.~{Serenelli}, F.L.~{Villante}, S.~{Basu}, J.~{Redondo} and
  J.~{Isern}, \emph{{New axion and hidden photon constraints from a solar data
  global fit}},
  \href{https://doi.org/10.1088/1475-7516/2015/10/015}{\emph{\jcap} {\bfseries
  2015} (2015) 015} [\href{https://arxiv.org/abs/1501.01639}{{\ttfamily
  1501.01639}}].

\bibitem{AN2015331}
H.~An, M.~Pospelov, J.~Pradler and A.~Ritz, \emph{Direct detection constraints
  on dark photon dark matter},
  \href{https://doi.org/https://doi.org/10.1016/j.physletb.2015.06.018}{\emph{Physics
  Letters B} {\bfseries 747} (2015) 331}.

\bibitem{Dolan:2022kul}
M.J.~Dolan, F.J.~Hiskens and R.R.~Volkas, \emph{{Advancing globular cluster
  constraints on the axion-photon coupling}},
  \href{https://doi.org/10.1088/1475-7516/2022/10/096}{\emph{JCAP} {\bfseries
  10} (2022) 096} [\href{https://arxiv.org/abs/2207.03102}{{\ttfamily
  2207.03102}}].

\bibitem{Redondo:2008aa}
J.~Redondo, \emph{{Helioscope Bounds on Hidden Sector Photons}},
  \href{https://doi.org/10.1088/1475-7516/2008/07/008}{\emph{JCAP} {\bfseries
  07} (2008) 008} [\href{https://arxiv.org/abs/0801.1527}{{\ttfamily
  0801.1527}}].

\bibitem{Chang:2016ntp}
J.H.~Chang, R.~Essig and S.D.~McDermott, \emph{{Revisiting Supernova 1987A
  Constraints on Dark Photons}},
  \href{https://doi.org/10.1007/JHEP01(2017)107}{\emph{JHEP} {\bfseries 01}
  (2017) 107} [\href{https://arxiv.org/abs/1611.03864}{{\ttfamily
  1611.03864}}].

\bibitem{Rrapaj:2019eam}
E.~Rrapaj, A.~Sieverding and Y.-Z.~Qian, \emph{{Rate of dark photon emission
  from electron positron annihilation in massive stars}},
  \href{https://doi.org/10.1103/PhysRevD.100.023009}{\emph{Phys. Rev. D}
  {\bfseries 100} (2019) 023009}
  [\href{https://arxiv.org/abs/1904.10567}{{\ttfamily 1904.10567}}].

\bibitem{Iglesias_1996}
C.A.~Iglesias and S.J.~Rose, \emph{Corrections to bremsstrahlung and thomson
  scattering at the solar center},
  \href{https://doi.org/10.1086/310172}{\emph{The Astrophysical Journal}
  {\bfseries 466} (1996) L115}.

\bibitem{1930RSPSA.126..654G}
J.A.~{Gaunt}, \emph{{Continuous Absorption}},
  \href{https://doi.org/10.1098/rspa.1930.0034}{\emph{Proceedings of the Royal
  Society of London Series A} {\bfseries 126} (1930) 654}.

\bibitem{Pradler2021}
J.~{Pradler} and L.~{Semmelrock}, \emph{{Accurate Gaunt Factors for
  Nonrelativistic Quadrupole Bremsstrahlung}},
  \href{https://doi.org/10.3847/1538-4357/ac0898}{\emph{\apj} {\bfseries 916}
  (2021) 105} [\href{https://arxiv.org/abs/2103.03248}{{\ttfamily
  2103.03248}}].

\bibitem{SommerfeldMaue}
A.~Sommerfeld and A.W.~Maue, \emph{Verfahren zur näherungsweisen anpassung
  einer lösung der schrödinger- an die diracgleichung},
  \href{https://doi.org/https://doi.org/10.1002/andp.19354140703}{\emph{Annalen
  der Physik} {\bfseries 414} (1935) 629}.

\bibitem{RevModPhys.34.507}
P.J.~Brussaard and H.C.~van~de Hulst, \emph{Approximation formulas for
  nonrelativistic bremsstrahlung and average gaunt factors for a maxwellian
  electron gas}, \href{https://doi.org/10.1103/RevModPhys.34.507}{\emph{Rev.
  Mod. Phys.} {\bfseries 34} (1962) 507}.

\bibitem{KL61}
W.J.~{Karzas} and R.~{Latter}, \emph{{Electron Radiative Transitions in a
  Coulomb Field.}}, \href{https://doi.org/10.1086/190063}{\emph{\apjs}
  {\bfseries 6} (1961) 167}.

\bibitem{RM-2580-AEC}
J.M.~Green, \emph{Fermi-Dirac Averages of the Free-Free Hydrogenic Gaunt
  Factor}, RAND Corporation, Santa Monica, CA (1960),
  \href{https://doi.org/10.7249/RM2580}{10.7249/RM2580}.

\bibitem{1987ApJS...63..661N}
M.~{Nakagawa}, Y.~{Kohyama} and N.~{Itoh}, \emph{{Relativistic Free-free Gaunt
  Factor of the Dense High-Temperature Stellar Plasma}},
  \href{https://doi.org/10.1086/191177}{\emph{\apjs} {\bfseries 63} (1987)
  661}.

\bibitem{bekefi1966radiation}
G.~Bekefi, \emph{Radiation Processes in Plasmas}, Radiation Processes in
  Plasmas, Wiley (1966).

\bibitem{ARMSTRONG201461}
G.~Armstrong, J.~Colgan, D.~Kilcrease and N.~Magee, \emph{Ab initio calculation
  of the non-relativistic free–free gaunt factor incorporating plasma
  screening},
  \href{https://doi.org/https://doi.org/10.1016/j.hedp.2013.10.005}{\emph{High
  Energy Density Physics} {\bfseries 10} (2014) 61}.

\bibitem{BetheHeitler}
H.~{Bethe} and W.~{Heitler}, \emph{{On the Stopping of Fast Particles and on
  the Creation of Positive Electrons}},
  \href{https://doi.org/10.1098/rspa.1934.0140}{\emph{Proceedings of the Royal
  Society of London Series A} {\bfseries 146} (1934) 83}.

\bibitem{1958MNRAS.118..241G}
I.P.~{Grant}, \emph{{Calculation of Gaunt factors for free-free transitions
  near positive ions}},
  \href{https://doi.org/10.1093/mnras/118.3.241}{\emph{\mnras} {\bfseries 118}
  (1958) 241}.

\bibitem{KippenhahnWeigert}
R.~{Kippenhahn}, A.~{Weigert} and A.~{Weiss}, \emph{{Stellar Structure and
  Evolution}} (2012),
  \href{https://doi.org/10.1007/978-3-642-30304-3}{10.1007/978-3-642-30304-3}.

\bibitem{Sweigert1978}
A.V.~{Sweigart} and P.G.~{Gross}, \emph{{Evolutionary sequences for red giant
  stars.}}, \href{https://doi.org/10.1086/190506}{\emph{\apjs} {\bfseries 36}
  (1978) 405}.

\bibitem{Raffelt:1992pi}
G.~Raffelt and A.~Weiss, \emph{{Nonstandard neutrino interactions and the
  evolution of red giants}}, {\emph{Astron. Astrophys.} {\bfseries 264} (1992)
  536}.

\bibitem{Straniero:2020iyi}
O.~Straniero, C.~Pallanca, E.~Dalessandro, I.~Dominguez, F.R.~Ferraro,
  M.~Giannotti et~al., \emph{{The RGB tip of galactic globular clusters and the
  revision of the axion-electron coupling bound}},
  \href{https://doi.org/10.1051/0004-6361/202038775}{\emph{Astron. Astrophys.}
  {\bfseries 644} (2020) A166}
  [\href{https://arxiv.org/abs/2010.03833}{{\ttfamily 2010.03833}}].

\bibitem{1994NIMPA.346..306D}
G.~{D'Agostini}, \emph{{On the use of the covariance matrix to fit correlated
  data}}, \href{https://doi.org/10.1016/0168-9002(94)90719-6}{\emph{Nuclear
  Instruments and Methods in Physics Research A} {\bfseries 346} (1994) 306}.

\bibitem{2022MNRAS.514.3058S}
I.D.~{Saltas} and E.~{Tognelli}, \emph{{New calibrated models for the tip of
  the red giant branch luminosity and a thorough analysis of theoretical
  uncertainties}}, \href{https://doi.org/10.1093/mnras/stac1546}{\emph{\mnras}
  {\bfseries 514} (2022) 3058}
  [\href{https://arxiv.org/abs/2203.02499}{{\ttfamily 2203.02499}}].

\bibitem{Lattanzio2}
T.~{Constantino}, S.W.~{Campbell}, J.C.~{Lattanzio} and A.~{van Duijneveldt},
  \emph{{The treatment of mixing in core helium burning models - II.
  Constraints from cluster star counts}},
  \href{https://doi.org/10.1093/mnras/stv2939}{\emph{\mnras} {\bfseries 456}
  (2016) 3866} [\href{https://arxiv.org/abs/1512.04845}{{\ttfamily
  1512.04845}}].

\bibitem{Ayala:2014pea}
A.~Ayala, I.~Domínguez, M.~Giannotti, A.~Mirizzi and O.~Straniero,
  \emph{{Revisiting the bound on axion-photon coupling from Globular
  Clusters}}, \href{https://doi.org/10.1103/PhysRevLett.113.191302}{\emph{Phys.
  Rev. Lett.} {\bfseries 113} (2014) 191302}
  [\href{https://arxiv.org/abs/1406.6053}{{\ttfamily 1406.6053}}].

\bibitem{Lattanzio3}
T.~{Constantino}, S.W.~{Campbell} and J.C.~{Lattanzio}, \emph{{The treatment of
  mixing in core helium-burning models - III. Suppressing core breathing pulses
  with a new constraint on overshoot}},
  \href{https://doi.org/10.1093/mnras/stx2321}{\emph{\mnras} {\bfseries 472}
  (2017) 4900} [\href{https://arxiv.org/abs/1709.06381}{{\ttfamily
  1709.06381}}].

\bibitem{Foot:1990mn}
R.~Foot, \emph{{New Physics From Electric Charge Quantization?}},
  \href{https://doi.org/10.1142/S0217732391000543}{\emph{Mod. Phys. Lett. A}
  {\bfseries 6} (1991) 527}.

\bibitem{He:1990pn}
X.-G.~He, G.C.~Joshi, H.~Lew and R.R.~Volkas, \emph{{New-$Z'$ Phenomenology}},
  \href{https://doi.org/10.1103/PhysRevD.43.R22}{\emph{Phys. Rev. D} {\bfseries
  43} (1991) 22}.

\bibitem{He:1991qd}
X.-G.~He, G.C.~Joshi, H.~Lew and R.R.~Volkas, \emph{{Simplest $Z'$ model}},
  \href{https://doi.org/10.1103/PhysRevD.44.2118}{\emph{Phys. Rev. D}
  {\bfseries 44} (1991) 2118}.

\bibitem{Friedland}
A.~{Friedland}, M.~{Giannotti} and M.~{Wise}, \emph{{Constraining the
  Axion-Photon Coupling with Massive Stars}},
  \href{https://doi.org/10.1103/PhysRevLett.110.061101}{\emph{\prl} {\bfseries
  110} (2013) 061101} [\href{https://arxiv.org/abs/1210.1271}{{\ttfamily
  1210.1271}}].

\bibitem{Dolan2021}
M.J.~{Dolan}, F.J.~{Hiskens} and R.R.~{Volkas}, \emph{{Constraining axion-like
  particles using the white dwarf initial-final mass relation}},
  \href{https://doi.org/10.1088/1475-7516/2021/09/010}{\emph{\jcap} {\bfseries
  2021} (2021) 010} [\href{https://arxiv.org/abs/2102.00379}{{\ttfamily
  2102.00379}}].

\bibitem{Lattanzio1}
T.~{Constantino}, S.W.~{Campbell}, J.~{Christensen-Dalsgaard}, J.C.~{Lattanzio}
  and D.~{Stello}, \emph{{The treatment of mixing in core helium burning models
  - I. Implications for asteroseismology}},
  \href{https://doi.org/10.1093/mnras/stv1264}{\emph{\mnras} {\bfseries 452}
  (2015) 123} [\href{https://arxiv.org/abs/1506.01209}{{\ttfamily
  1506.01209}}].

\bibitem{Spruit2015}
H.C.~{Spruit}, \emph{{The growth of helium-burning cores}},
  \href{https://doi.org/10.1051/0004-6361/201527171}{\emph{\aap} {\bfseries
  582} (2015) L2} [\href{https://arxiv.org/abs/1509.00659}{{\ttfamily
  1509.00659}}].

\bibitem{MIST1}
J.~{Choi}, A.~{Dotter}, C.~{Conroy}, M.~{Cantiello}, B.~{Paxton} and
  B.D.~{Johnson}, \emph{{Mesa Isochrones and Stellar Tracks (MIST). I.
  Solar-scaled Models}},
  \href{https://doi.org/10.3847/0004-637X/823/2/102}{\emph{\apj} {\bfseries
  823} (2016) 102} [\href{https://arxiv.org/abs/1604.08592}{{\ttfamily
  1604.08592}}].

\end{thebibliography}\endgroup
\end{document}